\documentclass[12pt,letterpaper]{article}

\usepackage{natbib,hyperref}
\usepackage{setspace}
\usepackage[ left=1in, top=1in, right=1in, bottom=1in]{geometry}
\usepackage{graphicx,bm,colonequals,amsmath,amssymb,url,xcolor,bbm}
\usepackage{array,tabularx,multirow}
\usepackage{enumitem}
\usepackage[font={footnotesize}]{caption,subcaption}
\usepackage[utf8]{inputenc}
\usepackage{enumitem}
\usepackage{soul} 
\usepackage{placeins} 
\usepackage[normalem]{ulem}
\usepackage{makecell}
\usepackage[version=4]{mhchem}
\graphicspath{{plots/}}

\usepackage{xr}
\makeatletter
\newcommand*{\addFileDependency}[1]{
  \typeout{(#1)}
  \@addtofilelist{#1}
  \IfFileExists{#1}{}{\typeout{No file #1.}}
}
\makeatother
\newcommand*{\myexternaldocument}[1]{%
    \externaldocument{#1}%
    \addFileDependency{#1.tex}%
    \addFileDependency{#1.aux}%
}
\myexternaldocument{supplement}

\usepackage[ruled,vlined]{algorithm2e}
\SetKwInput{KwInput}{Input}
\usepackage{algorithmic}
\usepackage{array,framed}
\usepackage{float}

\bibpunct[, ]{(}{)}{;}{a}{,}{,}
   
\usepackage{amsthm}
\newtheoremstyle{propstyle} 
    {2mm}                    
    {1mm}                    
    {\itshape}                   
    {}                           
    {\scshape}                   
    {.}                          
    {.5em}                       
    {}  
\theoremstyle{propstyle}

\theoremstyle{propstyle}

\theoremstyle{propstyle}

\theoremstyle{propstyle}

\theoremstyle{propstyle}

\makeatletter
\renewcommand{\paragraph}{%
  \@startsection{paragraph}{4}%
  {\z@}{2ex \@plus 1ex \@minus .2ex}{-1em}%
  {\normalfont\normalsize\bfseries}%
}
\makeatother

\DeclareMathAlphabet\mathbfcal{OMS}{cmsy}{b}{n}

\newcommand{\bu}{\mathbf{u}}

\newcommand{\bd}{\mathbf{d}}
\newcommand{\bb}{\mathbf{b}}
\newcommand{\bs}{\mathbf{s}}

\newcommand{\bx}{\mathbf{x}}
\newcommand{\by}{\mathbf{y}}

\newcommand{\bz}{\mathbf{z}}

\newcommand{\bA}{\mathbf{A}}

\newcommand{\bW}{\mathbf{W}}

\newcommand{\bI}{\mathbf{I}}

\newcommand{\bU}{\mathbf{U}}
\newcommand{\bV}{\mathbf{V}}

\newcommand{\bX}{\mathbf{X}}

\newcommand{\bB}{\mathbf{B}}
\newcommand{\bC}{\mathbf{C}}

\newcommand{\bfzero}{\mathbf{0}}

\newcommand{\bfmu}{\bm{\mu}}

\newcommand{\bftheta}{\bm{\theta}}

\newcommand{\bfbeta}{\bm{\beta}}
\newcommand{\bfepsilon}{\bm{\epsilon}}

\newcommand{\bfSigma}{\bm{\Sigma}}

\newcommand{\E}{\mathbb{E}}

\newcommand{\diag}{diag}

\newcommand{\GP}{\mathcal{GP}}

\newcommand{\order}{\mathcal{O}}
\newcommand{\normal}{\mathcal{N}}
\newcommand{\kronecker}{\raisebox{1pt}{\ensuremath{\:\otimes\:}}}
\DeclareMathOperator*{\argmin}{arg\,min}
\DeclareMathOperator*{\argmax}{arg\,max}
\DeclareMathOperator*{\pen}{p}

\newcommand{\locs}{\mathcal{S}}

\newcommand{\dens}{f}
\newcommand{\adens}{\widehat{f}}

\title{Scalable penalized spatiotemporal land-use regression for ground-level nitrogen dioxide}

\author{Kyle P.\ Messier\footnote{National Toxicology Program, National Institute of Environmental Health Sciences}\and Matthias Katzfuss\footnote{Department of Statistics, Texas A\&M University}
}

\begin{document}

\maketitle

\begin{abstract}
Nitrogen dioxide (NO$_2$) is a primary constituent of traffic-related air pollution and has well established harmful environmental and human-health impacts. Knowledge of the spatiotemporal distribution of NO$_2$ is critical for exposure and risk assessment. A common approach for assessing air pollution exposure is linear regression involving spatially referenced covariates, known as land-use regression (LUR). We develop a scalable approach for simultaneous variable selection and estimation of LUR models with spatiotemporally correlated errors, by combining a general-Vecchia Gaussian-process approximation with a penalty on the LUR coefficients. In comparisons to existing methods using simulated data, our approach resulted in higher model-selection specificity and sensitivity and in better prediction in terms of calibration and sharpness, for a wide range of relevant settings. 
In our spatiotemporal analysis of daily, US-wide, ground-level NO$_2$ data, our approach was more accurate, and produced a sparser and more interpretable model. Our daily predictions elucidate spatiotemporal patterns of NO$_2$ concentrations across the United States, including significant variations between cities and intra-urban variation. Thus, our predictions will be useful for epidemiological and risk-assessment studies seeking daily, national-scale predictions, and they can be used in acute-outcome health-risk assessments. 
\end{abstract}


\section{Introduction}

Nitrogen dioxide (NO$_2$) is a primary constituent of traffic-related air pollution and has well established harmful environmental and human-health impacts \citep{epa2016integrated}. For example, exposure to NO$_2$ is associated with increased all-cause mortality \citep{hoek2013long}, myocardial infarction \citep{rosenlund2006long,rosenlund2009traffic}, coronary heart disease \citep{rosenlund2008traffic}, cardiovascular events \citep{Alexeeff2018}, asthma \citep{gauderman2005childhood}, autism spectrum disorders \citep{Volk2013}, and impaired neurological development and other neurological disorders \citep{Xu2016}. Additionally, atmospheric oxides of nitrogen, including NO$_2$, are precursors to  hazardous acid rain \citep{schindler1988effects}, tropospheric ozone \citep{epa1999nitrogen}, fine particulate matter (PM$_{2.5}$) \citep{epa1999nitrogen}, and can result in negative ecological \citep{schindler1988effects} and economic impacts \citep{mauzerall2005nox}.  

Knowledge of the spatiotemporal distribution of NO$_2$ is critical for assessing exposure and subsequent risks. A common approach for assessing exposure to outdoor air pollution is linear regression involving spatially referenced covariates, known as land-use regression (LUR). There are many strengths in current implementations of LUR models. First, is the ability to predict a variable of interest in space and time at unmonitored coordinates, including uncertainty quantification. Second, is the use of readily-available, large geospatial datasets such as satellite imagery and census information.  Third, is the elucidation and interpretation of coefficients that are possible with linear models, which allows for meaningful policy discussions around factors affecting the distribution of exposure and risk.

Assuming independent and identically distributed (iid) errors, LUR has been implemented for air-quality-exposure modeling of NO$_2$ \citep{Briggs1997,Hoek2008, Su2009, Novotny2011,Ross2013,Knibbs2014,Larkin2017,DeHoogh2018} and other air pollutants such as PM$_{2.5}$ \citep{Henderson2007, Moore2007,Ross2013}. Typically, LUR involves model selection or dimension reduction on a large candidate-set of spatially referenced covariates. For example, LUR has been implemented with stepwise model selection for NO$_2$ \citep{Briggs1997,Su2009, Novotny2011,Ross2013,Knibbs2014,DeHoogh2018} and partial-least-squares dimension reduction for NO$_2$ \citep{Young2016} and PM$_{2.5}$ \citep{Sampson2013}. NO$_2$ LUR models have also employed penalization-based model-selection methods such as the LASSO \citep{Knibbs2014,Larkin2017}. Additionally, LUR prediction residuals are often integrated into geostatistical models, such as Kriging \citep{Wu2013,DeHoogh2018} and Bayesian maximum entropy \citep{Coulliette2009, Messier2012, Beckerman2013, Reyes2014, Messier2014, Messier2015}, in a two-stage approach with the goal of improving prediction accuracy. 

While LURs have undoubtedly been useful for many exposure and risk assessment studies, the assumption of iid errors is usually violated, because the spatial dependence in the response cannot be captured fully by the covariates, resulting in biased covariate estimates and decreased sensitivity and specificity in the model-selection process. An exception to this case is \citet{Holcomb2018}, which implemented backwards model selection in a full Kriging model, 
but this approach is not feasible for large data sets. \citet{Guan2020} implemented a scalable approach with LUR with spatiotemporal errors, but used principal components instead of model selection to reduce the number of covariates.

In spatial statistics and Gaussian-process modeling, many approaches have been proposed to ensure scalability to large datasets \citep[see, e.g.,][for recent reviews and comparison]{Heaton2017, Liu2020}, but the focus is often more on prediction based on the (residual) covariance structure, and less on penalized selection from among a large number of spatial or spatiotemporal covariates. Perhaps the most promising approaches for scalable spatial prediction are based on the ordered conditional approximation of \citet{Vecchia1988}; here we use and extend the general Vecchia approximation \citep{Katzfuss2017a,Katzfuss2018}, which is highly accurate, can guarantee linear complexity with respect to the sample size, and includes many existing Gaussian-process approximations as special cases \citep[e.g.,][]{Vecchia1988,Snelson2007,Finley2009,Sang2011a,Datta2016,Katzfuss2015,Katzfuss2017b}.

We develop an approach for simultaneous variable selection and estimation of LUR models with spatiotemporally correlated errors, extending the general Vecchia approximation to ensure scalability to large datasets. The resulting dependent-error regression problem can be transformed into standard iid-error regression involving pseudo-data, which can be computed rapidly using Vecchia. This approach can be combined with any existing method for fitting penalized regression models with independent errors, such as least-angle regression \citep{Efron2004} for LASSO-type L1 penalties \citep{Tibshirani1996}, and coordinate descent \citep{breheny2011} for non-convex (e.g., smoothly clipped absolute deviation) penalties \citep{FanLi2001}. The ordering and conditioning-set selection necessary for the Vecchia approximation is carried out based on appropriately scaled spatiotemporal coordinates. All computations necessary for inference scale linearly in the data size for fixed tuning parameters.

The remainder of this article is organized as follows: Section \ref{sec:data} describes the daily, ground-level NO$_2$ data and the geographic covariates. Section \ref{sec:lur} provides a description of LUR with penalization. Section \ref{sec:methodology} presents our proposed methodology based on the general Vecchia approximation to Kriging models with SCAD penalty. Section \ref{sec:simulation} compares approaches in simulation studies. In Section \ref{sec:application}, we apply our method to the NO$_2$ concentrations and discuss the results. Section \ref{sec:conclusions} highlights the main conclusions and discusses areas for future research.
Further details can be found in Appendices \ref{app:vecchia}--\ref{app:objective}. A separate Supplementary Material document contains Sections \ref{Sec:Joint-Varying}--\ref{sec:cov-comp} with additional plots and details. Code based on the \texttt{R} package \texttt{GPvecchia} \citep{GPvecchia} for the proposed approach and this study is available at \url{https://github.com/NIEHS/LURK-Vecchia}.

\section{Ground-level NO$_2$ data\label{sec:data}}

\begin{figure}
\centering\includegraphics[width =.5\linewidth]{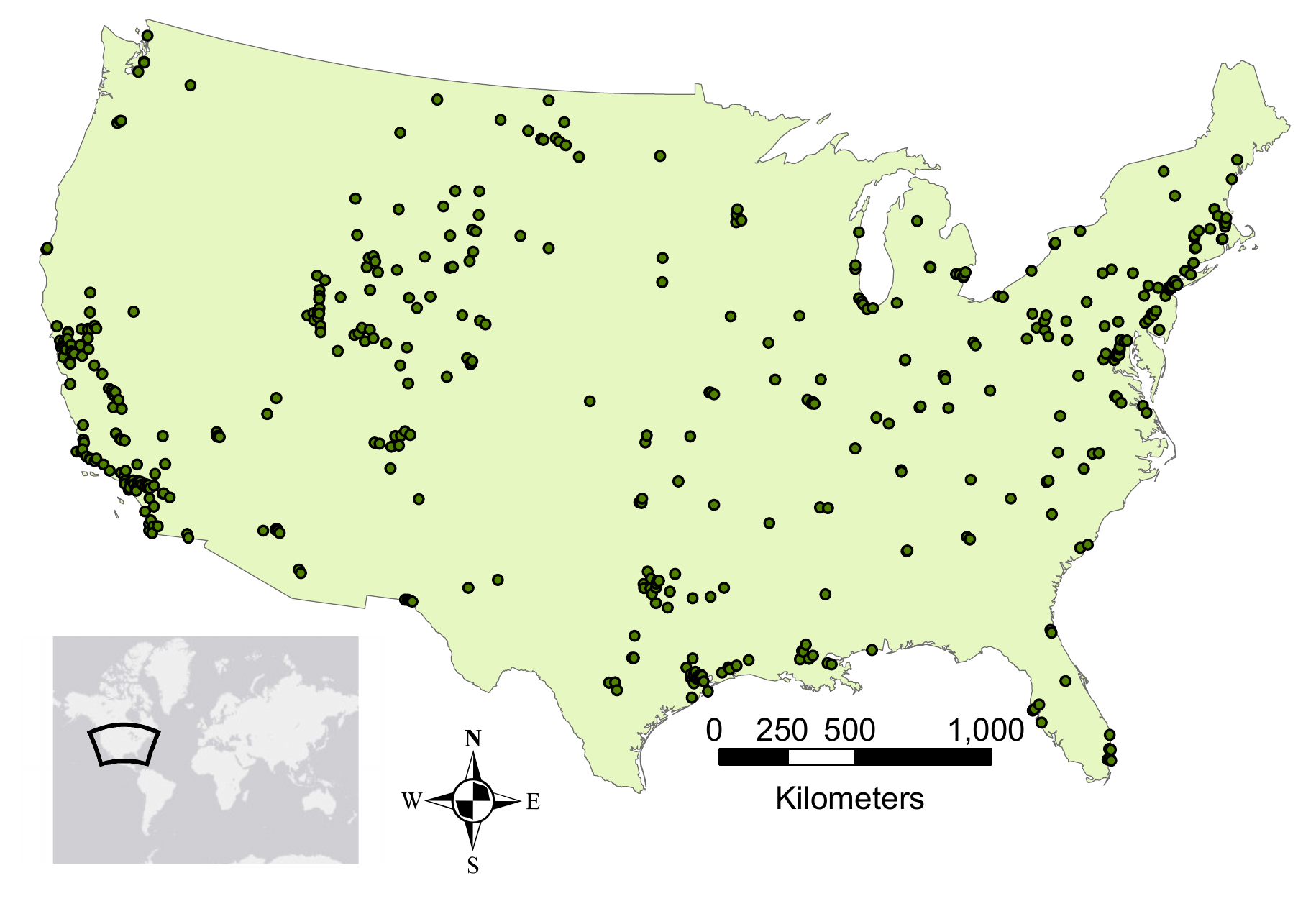}
  \caption{Locations of the 459 monitoring sites for the USEPA AQS NO$_2$ data over the study domain (conterminous United States)}
\label{fig:Domain}
\end{figure}

We consider daily ground-level (i.e., tropospheric) NO$_2$ concentrations across the conterminous United States, monitored and distributed by the United States Environmental Protection Agency (USEPA) Air Quality System (AQS) \citep{epa2019data}. The date range for our study was July 10, 2018 to May 1, 2019, based on the availability of geographic covariates, primarily the TROPOMI real-time satellite imagery. The final NO$_2$ dataset contained 76,748 unique spatiotemporal observations distributed across 459 monitoring sites (Figure \ref{fig:Domain}).

\subsection{Geographic covariates}\label{s:covariates}

For our analysis, we calculated 139 spatial and spatiotemporal geographic covariates representing possible NO$_2$ sources and attenuation factors. A key characteristic of our and the majority of LUR studies is the presence of highly correlated covariates. In particular, many covariates only differ by their spatial resolution. After all of the covariates are calculated, each covariate is standardized to mean 0 and variance 1. The following subsections explain how each potential covariate was calculated.

\subsubsection{TROPOMI}\label{sec:TROPOMI}

We utilized data from the TROPOspheric Monitoring Instrument (TROPOMI) to calculate many satellite-based spatiotemporal covariates. TROPOMI is the sensor on-board the Copernicus Sentinel-5 Precursor satellite. The TROPOMI-based covariate calculations are performed in Google Earth Engine, a cloud platform for earth observation data analysis that combines a public data catalog with a large-scale computational facility optimized for parallel processing of geospatial data. 

TROPOMI provides output (i.e., Level-2 or L2 products) representing atmospheric air pollution and physical properties with a spatial resolution of approximately 3.5 by 7 km. We calculated daily mean values within 1, 10, and 100 km circular buffers for the following TROPOMI L2 products. Note the 1 km buffer is equivalent to the coincident TROPOMI value at the location of the monitor: Total vertical column NO$_2$ (mol-m$^{-2}$), tropospheric vertical column NO$_2$ (mol-m$^{-2}$), NO$_2$ slant column density (mol-m$^{-2}$), tropopause (i.e. boundary between troposphere and stratosphere) pressue (Pa), absorbing aerosol index (AAI; dimensionless), cloud fraction, and the solar azimuth angle (degrees). We used the near real-time TROPOMI product, if available, for the estimation of models. This was the driving factor for the sparsity of the data set within the study range. For prediction, if the real-time data were unavailable, we used the offline data. The near real-time data are available sooner and have small differences with the offline data \citep{boersma2007}. Lastly, we used a simple average from 10 nearest-neighbor spatiotemporal coordinates if neither near real-time nor offline were observed at a spatiotemporal coordinate. Potential alternatives for interpolating missing TROPOMI data include longer-time-scale moving-window averages (e.g. monthly) or developing a predictive model \citep{de2019predicting}. The final covariate dataset included 21 TROPOMI-based variables.

\subsubsection{Meteorology}\label{sec:Meteorology}

Spatial and daily time-resolved meteorological covariates were calculated in Google Earth Engine using the University of Idaho Gridded Surface Meteorological dataset (GRIDMET) \citep{abatzoglou2014}. GRIDMET provides daily surface fields at approximately 4 km resolution. We calculated average daily values inside 1, 10, and 100 km buffers for the following variables: precipitation (mm), maximum relative humidity (percent), specific humidity (kg-kg$^{-1}$), surface downward shortwave radiation (W-m$^{-2}$), maximum temperature (K), and wind velocity (m-s$^{-1}$). The 1 km buffer is equivalent to choosing the containing grid cell. The final covariate dataset included 18 meteorology-based variables. 

\subsubsection{Vegetative indices}\label{sec:veg}

Spatial covariates of vegetative indices were calculated in Google Earth Engine using the MODIS/Terra Vegetative Indices 16-Day L3 Global 500 m SIN Grid \citep{aster2019}. We calculated spatial averages of the normalized difference vegetative index (NDVI) and the enhanced vegative index (EVI) in 1, 10, and 100 km circular buffers. The final covariate dataset included 6 vegetative index based variables.

\subsubsection{Population, traffic, and roads}\label{sec:Population}

Population density (people-km$^{-2}$) was calculated in Google Earth Engine from the Gridded Population of World Version 4 \citep{GPWv4}. Average population (2015-equivalent) density was calculated in 1, 10, and 100 km circular buffers. 

A surrogate for traffic was calculated using the University of Oxford Malaria Atlas Project global travel friction dataset \citep{weiss2018}. Average travel friction  (min-m$^{-1}$), or travel time, was calculated in 1, 10, and 100 km circular buffers. 

Road length variables were calculated in ArcMap 10.6.1 and MATLAB R2018a using the ESRI major roads shapefile (Esri, TomTom North America, Inc). The road length in 1, 10, and 100 km circular buffers was calculated for the following road classifications (FRC code in ESRI shapefile): All roads classes, highway (0), major roads (1,2), and secondary roads (3,4,5). The final covariate dataset included 18 population, traffic, or road variables.

\subsubsection{Land cover}\label{sec:landcover}

Spatial land-cover attributes were calculated in Google Earth Engine from the National Land Cover Database \citep{homer2015}. The percent of each land cover class (e.g. water, low developed, deciduous trees, etc.) were calculated in 1, 10, and 100 km circular buffers. 

Average elevation was calculated in 1, 10, and 100 km circular buffers using the Japan Aerospace Exploration Agency Advanced Land Observing Satellite global digital surface model with a horizontal resolution of approximately 30 meters \citep{tadono2014}. The final covariate dataset included 48 land-cover or elevation variables.

\subsubsection{National Emissions Inventory}\label{sec:NEI}

 Data on point source emissions was downloaded from the USEPA National Emissions Inventory \citep{nei2017} for the year 2017. Following \citet{Messier2012}, we calculated NO$_2$ point source emissions as the sum of isotropic, exponentially decaying contributions from the point sources. The initial value was the total 2017 NO$_2$ emissions from the emissions inventory and decay ranges were a series of ranges from short to long distance decay ranges: 1 to 10 km by 1 km increments; 20 to 100 km by 10 km increments; and 200 to 1000 km by 100 km increments, resulting in 28 NEI-based covariates.

\section{Land-use regression with penalization\label{sec:lur}}

Let $\bz = (z_1,\ldots,z_n)'$ denote the response vector, where $z_i=z(\bs_i,t_i)$ is the log-transformed NO$_2$ measured on day $t_i$ at spatial location $\bs_i$. (We denote the response by $\bz$ here for consistency with papers on general Vecchia.) We have the values $\bx_i = \bx(\bs_i,t_i) = (x_1(\bs_i,t_i),\ldots,x_p(\bs_i,t_i))'$ of the $p=139$ covariates (described in Section \ref{s:covariates}) at the same (space, time)-coordinate pairs $\locs=\{(\bs_1,t_1),\ldots,(\bs_n,t_n)\}$. Spatial-only covariates are repeated in time as needed.  We assume a linear relationship between the response and covariates,
\begin{equation}
\label{eq:regressioni}
z_i = \bx_i'\bfbeta + \epsilon_i = \bx_i'\bfbeta + \eta_i + \delta_i = y_i + \delta_i, \qquad i =1,\ldots,n,
\end{equation}
where $\epsilon_i = \eta_i + \delta_i$ is the regression error consisting of a spatiotemporally dependent component $\eta_i=\eta(\bs_i,t_i)$ and an independent measurement-noise component $\delta_i \stackrel{iid}{\sim} N(0,\tau^2)$, $i=1,\ldots,n$, and $y_i= z_i -\delta_i = \bx_i'\bfbeta + \eta_i$ is the (noise-free) true log-NO$_2$. We assume that $\eta(\cdot) \sim \GP(0,C_{\bftheta})$ follows a Gaussian process with covariance function $C_{\bftheta}$. Throughout, we assume a non-separable spatiotemporal exponential covariance function, $C_{\bftheta}\big((\bs_i,t_i),(\bs_j,t_j)\big) = \sigma^2 \text{exp}\big(-d_{\bftheta}\big((\bs_i,t_i),(\bs_j,t_j)\big)\big)$, where
  \begin{equation}
  \label{eq:dist}
 \textstyle d_{\bftheta}\big((\bs_i,t_i),(\bs_j,t_j)\big) = \sqrt{ \frac{||\bs_i - \bs_j||_2^2}{\gamma_s^2} +  \frac{(t_i - t_j)^2}{\gamma_t^2}},
  \end{equation}
and $\bftheta = (\sigma,\gamma_s,\gamma_t,\tau)$ contains the unknown parameters in the model. We also considered and dismissed a Mat\'ern covariance with estimated smoothness parameter, as this resulted in a smoothness parameter near that of the exponential model (i.e., 0.5), nearly identical negative loglikelihood values, and increased model run time due to evaluation of Bessel functions and larger parameter space. Note that standard LUR models \citep[e.g.,][]{Briggs1997, Hoek2008, DeHoogh2018} do not include the dependent component $\eta(\cdot)$ and assume the error terms $\epsilon_1,\ldots,\epsilon_n$ to be iid.

Stacking the quantities in \eqref{eq:regressioni}, we obtain the regression model
\[
 \bz = \bX \bfbeta + \bfepsilon, \qquad \bfepsilon \sim \normal_n(\bfzero,\bfSigma_{\bftheta}),
\]
where $\bfSigma_{\bftheta} = \bC_{\bftheta} + \tau^2 \bI_n$, with $\bC_{\bftheta} = \big(C_{\bftheta}((\bs_i,t_i),(\bs_j,t_j))\big)_{i,j=1,\ldots,n}$.
Equivalently, we can write this in terms of a multivariate Gaussian density for the response,
\begin{equation}
\label{eq:lik}
 f(\bz;\bfbeta,\bftheta) = \normal_n(\bz| \bX\bfbeta,\bfSigma_{\bftheta}).
\end{equation}

The goal is to estimate the $p$-vector $\bfbeta$ and determine its nonzero elements, which also requires estimation of the covariance parameters $\bftheta$. Further, given parameter estimates $\hat\bfbeta$ and $\hat\bftheta$, we would like to predict the process $y(\cdot)$ at unobserved coordinates. 

A standard approach for parameter estimation is to maximize the likelihood in \eqref{eq:lik} with respect to the parameters $\bfbeta$ and $\bftheta$. However, we have a large number $p=139$ of (correlated) covariates, which makes the least-squares or maximum likelihood estimates of $\bfbeta$ unstable. To alleviate this issue, and to be able to select certain variables and set the coefficients corresponding to the other variables to zero, we instead consider optimizing an objective function consisting of the negative loglikelihood plus a penalization term $\pen(\bfbeta)$ on $\bfbeta$: 
\begin{equation}
\label{eq:objective}
Q(\bfbeta,\bftheta) = -2 \log f(\bz;\bfbeta,\bftheta) + \lambda\, \pen(\bfbeta) = (\bz - \bX\bfbeta)'\bfSigma_{\bftheta}^{-1}(\bz - \bX\bfbeta) + \log |\bfSigma_{\bftheta} | + \lambda \, \pen(\bfbeta),
\end{equation}
where $\lambda$ is a shrinkage or tuning parameter, and we have omitted an additive constant in the last equation. In our numerical examples and application, we will use the popular non-convex, smoothly clipped absolute deviation (SCAD) penalty \citep{FanLi2001}: 
\begin{equation}
\label{eq:SCAD}
  \text{p}(\beta) =
    \begin{cases}
      \lambda |{\beta}|, & \text{if}\hspace{2mm}  |{\beta}| \leq \lambda, \\
      \frac{2a\lambda |\beta| - \beta^2 - \lambda^2}{2(a-1)}, & \text{if} \hspace{2mm} \lambda <  |\beta| \leq a\lambda,\\
      \frac{\lambda^2 (a+1)}{2}, & \text{otherwise},
    \end{cases}   
\end{equation}
where $a = 3.7$, a popular choice that performs comparably to values based on generalized cross-validation \citep{FanLi2001}. We use the SCAD penalty for its oracle property, but other penalties can be easily be swapped in our framework.  \citet{Li2005} demonstrate that a SCAD-penalized likelihood as in \eqref{eq:objective} and \eqref{eq:SCAD} reduces the variance in the estimates of $\bftheta$; however, their discussion did not address model selection or large sample sizes.

\section{Our methodology \label{sec:methodology}}

\subsection{A general Vecchia approximation of the objective function}

Evaluation or optimization of the objective function $Q$ in \eqref{eq:objective} requires decomposition of the $n\times n$ covariance matrix $\bfSigma_{\bftheta}$ for many different values of $\bftheta$, each of which takes $\order(n^3)$ time. This is computationally infeasible for the large $n=76{,}748$ in our application.

Hence, we will extend the sparse general Vecchia (SGV) approximation \citep{Katzfuss2017a}, which we briefly review here, with more details given in Appendix \ref{app:vecchia}. SGV applies the approximation of \citet{Vecchia1988} to the vector $\bu=(y_1,z_1,\ldots,y_n,z_n)'$, which interweaves the latent true-process realizations $y_1,\ldots,y_n$ and the observed noisy data $z_1,\ldots,z_n$. This approximation essentially replaces the conditioning sets in the exact factorization $\dens(\bu) = \prod_{j=1}^{2n} \dens(u_j|u_1,\ldots,u_{j-1})$ by small subsets, resulting in the approximation
\begin{equation}
\label{eq:vecchiaapprox}
\adens(\bu)  = \prod_{i=1}^{2n} p(u_i|\bu_{g(i)}) = \normal_{2n}\big(\bu|(\bX \kronecker \mathbf{1}_2)\bfbeta,(\bU\bU')^{-1}\big)
\end{equation}
where each $g(i) \subset (1,\ldots,i-1)$ is a conditioning index set of size $|g(i)|\leq m$, $\kronecker$ is the Kronecker product, $\mathbf{1}_2$ is a vector consisting of two $1$s, and $\bU=\bU_{\bftheta}$ is a sparse upper triangular matrix whose nonzero entries can be computed easily based on $C_{\bftheta}$ and $\tau^2$.
Recent results \citep{Schafer2020} indicate that the approximation error can be bounded with the conditioning-set size $m$ increasing logarithmically in $n$ in some settings; in practice, $m \approx 30$ is often sufficient for accurate approximations.
We further define $\bA$ and $\bB$ as the submatrices of $\bU$ consisting of the odd- and even-numbered rows of $\bU$, corresponding to $\by$ and $\bz$, respectively. Then, $\bW = \bA\bA'$ is the implied posterior precision matrix of $\by$ given $\bz$, and we define $\bV$ as the Cholesky factor based on reverse row-column ordering of $\bW$.

Our approximation $\adens(\bu)$ is an extension of the SGV approach for spatial processes described in \citet{Katzfuss2017a}; to approximate the spatiotemporal covariance function $C_{\bftheta}$, we modify the ordering and conditioning scheme here to be carried out based on the scaled spatiotemporal distance \eqref{eq:dist}, which depends on unknown parameters and must be updated along with the parameters. Again, more details are given in Appendix \ref{app:vecchia}.

The SGV approximation of the density of $\bu=(y_1,z_1,\ldots,y_n,z_n)'$ in \eqref{eq:vecchiaapprox} implies an approximation of the distribution for the response:
\[
\adens(\bz) = \int \adens(\bu) d\by,
\]
which is also multivariate normal. This concludes our review of \citet{Katzfuss2017a}. Plugging the approximation $\adens(\bz)$ into \eqref{eq:objective} results in the Vecchia objective function
\begin{equation}
\label{eq:vobjective}
\hat Q(\bfbeta,\bftheta) = -2 \log \adens_{\bfbeta,\bftheta}(\bz) + \lambda\, \pen(\bfbeta),
\end{equation}
where we have now made explicit the dependence of the distribution of $\bz$ on the parameters $\bfbeta$ and $\bftheta$. We will use and optimize the Vecchia objective function $\hat Q(\bfbeta,\bftheta)$ in the remainder of the manuscript.

\subsection{Inference}

The most straightforward way to optimize the objective function is to optimize iteratively with respect to $\bftheta$ and $\bfbeta$, while holding the respective other parameter vector fixed. In practice, it is usually sufficient to do this just a small number of times, after which there is little change in the parameter values.

As we prove in Appendix \ref{app:objective}, $\hat Q(\bfbeta,\bftheta)$ in \eqref{eq:vobjective} can be written as
\begin{equation}
\label{eq:qvecchia}
\hat Q(\bfbeta,\bftheta) = \|\tilde\bz_{\bftheta} - \tilde\bX_{\bftheta}\bfbeta\|_2^2 + \lambda\, \pen(\bfbeta) - 2 \sum_{i} \log \big((\bU_{\bftheta})_{ii}\big) + 2\sum_{i} \log \big((\bV_{\bftheta})_{ii}\big),
\end{equation}
where $\tilde\bz_{\bftheta} = \bB'\bz + \bA'\bW^{-1} \bA\bB'\bz$, $\tilde\bX_{\bftheta} = \bB'\bX + \bA'\bW^{-1}\bA\bB'\bX$, and $\bV_{\bftheta}$ depend on $\bftheta$ through the matrices $\bA$, $\bB$, and $\bW$ computed from $\bU=\bU_{\bftheta}$.

\subsubsection{Estimation of the regression coefficients\label{sec:betaest}}

Optimizing $\hat Q(\bfbeta,\bftheta)$ in \eqref{eq:qvecchia} with respect to $\bfbeta$ for fixed $\bftheta=\hat\bftheta$ is equivalent to solving a standard penalized regression problem with iid errors, except based on the pseudo-data $\tilde\bz_{\hat\bftheta}$ and $\tilde\bX_{\hat\bftheta}$.

To develop some intuition, consider briefly the case $m=n-1$, in which case the approximation $\adens(\bu)$ in \eqref{eq:vecchiaapprox} becomes exact. Then, the pseudo-data $\tilde\bz_{\hat\bftheta}$ are obtained by first creating an augmented data vector of length $2n$ consisting of $\bz$ and $\E(\by|\bz)$, and then transforming this vector to a vector of iid normal variables based on the joint distribution or covariance matrix of $\bz$ and $\by$.
Interestingly, the resulting inference on $\bfbeta$ is unchanged relative to simply transforming the data $\bz$ alone, as is often done for general linear models. In the case of small $m$, our sparse general Vecchia approximation allows us to carry out this inference on $\bfbeta$ more accurately based on the first approach.

For example, in the case of the SCAD \citep{FanLi2001} or L1 penalties \citep{Tibshirani1996}, solution paths of optimal $\bfbeta$ values for each value of $\lambda$ can be computed rapidly using coordinate descent \citep{breheny2011} or least angle regression \citep{Efron2004}, respectively. We select the optimal $\lambda$ and the corresponding $\hat\bfbeta$ based on the lowest cross-validated mean square error, which can be performed in many software packages such as \textit{ncvreg} \citep{breheny2011} or \textit{glmnet} \citep{Friedman2010}. \citet{breheny2011} demonstrated for high-dimensional problems that the SCAD penalty estimated with coordinate descent, in combination with cross-validation, leads to the global minimum solution as it likely resides in the locally convex region of $\lambda$.

\begin{figure}
\centering\includegraphics[width =.5\linewidth]{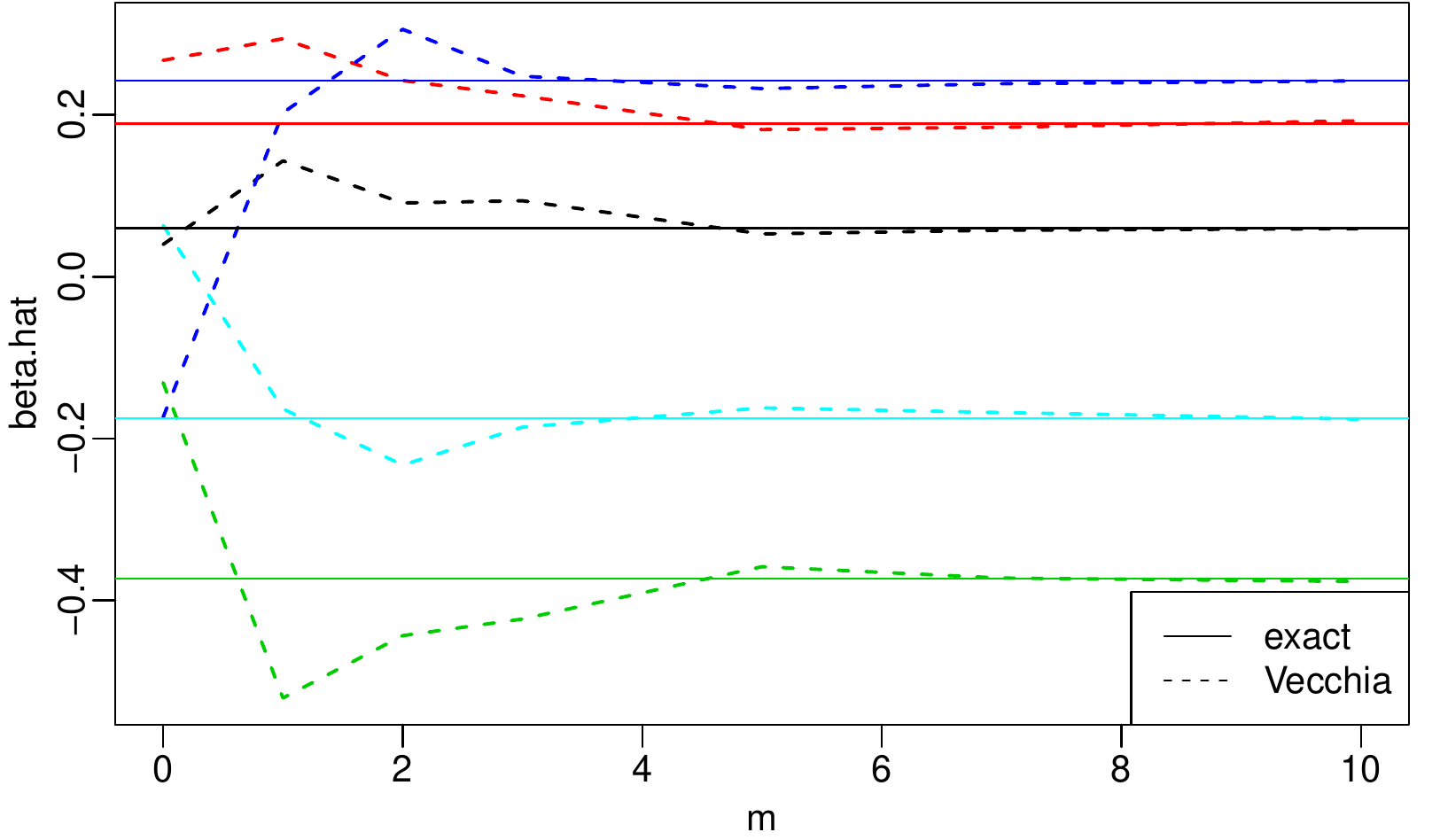}
  \caption{Demonstration of the quick convergence of general Vecchia estimates of $\bfbeta$ with increasing $m$ towards the exact (i.e., without approximation) generalized-least squares (GLS) estimate}
\label{fig:KL1D}
\end{figure}

We considered a simple example to illustrate how quickly the Vecchia solution can converge to the exact solution in the estimation of trend parameters, $\bfbeta$. We simulated $p=5$ correlated covariates with correlations ranging from $.45$ to $.82$, set the true $\bfbeta = \bfzero$, and simulated data with spatially dependent error at $n=500$ locations. We then computed the exact generalized least-squares (GLS) estimates and the Vecchia GLS estimates implied by the pseudo-data in \eqref{eq:qvecchia}. As shown in Figure \ref{fig:KL1D}, the Vecchia solution quickly approached the exact solution going from $m=0$ (i.e., assuming independent errors) to $m=10$.

\subsubsection{Estimation of the covariance parameters}

Defining $\bfepsilon_{\bfbeta} = \bz - \bX\bfbeta$, \eqref{eq:qvecchia} can be rearranged to yield
\begin{equation}
\label{eq:likelihood}
\hat Q(\bfbeta,\bftheta) = \|\bB'\bfepsilon_{\bfbeta}\|_2^2 - \|\bV^{-1}\bA \bB'\bfepsilon_{\bfbeta}\|_2^2 - 2 \sum_{i} \log \bU_{ii} + 2\sum_{i} \log \bV_{ii} + \lambda\, \pen(\bfbeta),
\end{equation}
where $\bV$, $\bA$, and $\bB$ implicitly depend on $\bftheta$ through $\bU= \bU_{\bftheta}$. Note that this expression is an extension of the Vecchia loglikelihood in \citet{Katzfuss2017a}; we replaced their zero-mean data $\bz$ with our residuals $\bfepsilon_{\bfbeta}$, and we have added the penalization term $\lambda\, \pen(\bfbeta)$. This alternative expression of the objective function is important, as it avoids having to compute the pseudo-data for every evaluation as in \eqref{eq:qvecchia}.

For a given $\bfbeta = \hat\bfbeta$, \eqref{eq:likelihood} can be evaluated cheaply for any given parameter value $\bftheta$, and hence we can optimize the objective function with respect to $\bftheta$ using standard numerical optimization algorithms (e.g., Nelder-Mead). As in most Gaussian-process models, there is no guarantee that this procedure will find the global optimum, but we have not observed any negative consequences.
Similarly, we have not observed any numerical issues due to nonidentifiability between the variance and range parameters, which theoretically holds under in-fill asymptotics \citep{Zhang2004,Tang2019} but not under the increasing-domain asymptotics that may be more appropriate for our real-data application with small effective ranges relative to the domain size.

We monitor convergence of the overall algorithm by considering the minimum value of \eqref{eq:likelihood} achieved at each iteration.

\subsubsection{Prediction \label{sec:pred}}

Often, interest is in prediction of the noise-free process $\by_P$ at a set of $n_P$ spatiotemporal coordinates $\locs_P$, which is equivalent to obtaining the conditional distribution of $\by_P$ given the data $\bz$. To do so, we extend the response-first full-conditioning (RF-full) approach of \citet{Katzfuss2018}, which essentially consists of a general Vecchia approximation $\adens(\tilde\bu)$, similar to \eqref{eq:vecchiaapprox}, but now applied to the vector $\tilde\bu=(\bz',\by_{\text{all}}')'$, where $\by_{\text{all}} = (\by',\by_P')'$. We have 
\[
\adens( \by_{\text{all}} | \bz ) = \frac{\adens(\bz,\by_{\text{all}})}{\int \adens(\bz,\by_{\text{all}})d\by_{\text{all}}} \equalscolon \normal_{n}(\bfmu_{\text{all}},\bW_{\text{all}}^{-1}).
\]
Any quantities of interest can be extracted from this \emph{joint} distribution, after computing $\bV_{\text{all}}$ as the Cholesky factor based on reverse row-column ordering of $\bW_{\text{all}}$. For example, the prediction mean, also referred to as the kriging predictor, is obtained by subsetting the vector $\bfmu_{\textnormal{all}} = \bX_{\textnormal{all}}\hat\bfbeta -(\bV_{\text{all}}')^{-1}\bV_{\text{all}}^{-1}\bA_{\textnormal{all}} \bB_{\textnormal{all}}'(\bz - \bX\hat\bfbeta) = (\bfmu',\bfmu_P')'$, where $\bX_{\textnormal{all}} = (\bX',\bX_P')'$, while the prediction or kriging variances are given by a subset of the diagonal elements of $\bW_{\textnormal{all}}^{-1}$, which can be obtained using selected inversion based on the Takahashi recursions for $\bV_{\text{all}}$.
Note that we ignore uncertainty in $\hat\bfbeta$ in the predictions, but we conducted experiments that showed this uncertainty is often small relative to the uncertainty in $\eta(\cdot)$.
If prediction of $\bz_P$ is desired, we simply need to add $\tau^2$ to the prediction variances. Our approximation is an extension of the spatial RF-full prediction in \citet{Katzfuss2018}, in that the nonzero mean has to be added and subtracted in the kriging predictor, and we carry out the ordering and conditioning in the scaled spatiotemporal domain. Details are given in Appendix \ref{app:vecchia}.

\begin{algorithm}[t]
\caption{LURK-Vecchia: Land-use regression Kriging with Vecchia approx.}
\KwInput{$\bz$, $\locs$, $\bX$, $\locs_P$, $\bX_P$, $C_{\bftheta}$, \texttt{tol}}
\KwResult{Parameter estimates $\hat\bfbeta$ and $\hat\bftheta$; prediction $\adens( \by_{\text{all}} | \bz )$ }
\begin{algorithmic}[1]
\STATE Initialize $\hat\bfbeta$, $\hat\bftheta$, \texttt{prev.objective} = $\infty$, and \texttt{converged} = \texttt{FALSE}
\STATE \texttt{OC}: Maxmin ordering and nearest-neighbor conditioning for $\locs$ based on $d_{\hat\bftheta}$ in \eqref{eq:dist} 
\WHILE{\texttt{converged = FALSE}}
 \STATE Compute $\hat\bftheta = \argmax_{\bftheta} \hat Q(\hat\bfbeta,\bftheta)$ based on \eqref{eq:likelihood} using \texttt{OC}
 \STATE Update \texttt{OC} based on $d_{\hat\bftheta}$ in \eqref{eq:dist} \label{l:ocupdate}
 \STATE Compute pseudo-data $\tilde\bz_{\hat\bftheta}$ and $\tilde\bX_{\hat\bftheta}$ as in \eqref{eq:qvecchia} using \texttt{OC}
 \STATE Estimate $\bfbeta$ using standard (iid) SCAD based on $\tilde\bz_{\hat\bftheta}$ and $\tilde\bX_{\hat\bftheta}$ (see Section \ref{sec:betaest}) \label{l:scad}
 \STATE \texttt{new.objective} = $\hat Q(\hat\bfbeta,\hat \bftheta)$
 \IF{\texttt{new.objective} $>$ $($\texttt{prev.objective} $\times$ (1-\texttt{tol}) $)$ }
  \STATE \texttt{converged} = \texttt{TRUE}
 \ELSE \STATE \texttt{prev.objective} = \texttt{new.objective}
 \ENDIF
\ENDWHILE
\STATE \texttt{OC}$_P$: Ordering and conditioning for $\locs$ and $\locs_P$ based on $d_{\hat\bftheta}$ (see Appendix \ref{app:vecchia})
\STATE Prediction: Compute relevant summaries of $\adens( \by_{\text{all}} | \bz )$ using \texttt{OC}$_P$ (see Section \ref{sec:pred})
\end{algorithmic}
\label{alg:lurkv}
\end{algorithm}

\subsubsection{Complexity}

Our proposed inference procedure is summarized in Algorithm \ref{alg:lurkv}.
If each conditioning index vector in our Vecchia approximations is at most of size $m$, SGV and RF-full ensure that $\bU$, $\bV$, and $\bV_{\text{all}}$ are all highly sparse with at most $m$ nonzero off-diagonal entries per column. As a consequence, for fixed $m$ and $p$, our entire inference procedure requires linear time in the number of observed and prediction coordinates.

More precisely, assuming that $m,p,n_P \leq n$, evaluation of the likelihood and prediction for each parameter value requires $\order(nm^3)$ time, 
coordinate descent for SCAD requires $\order(np)$ time per iteration, one triangular solve involving $\bV$ requires $\order(nm)$ time, and hence computing the pseudo-data $\tilde\bz_{\hat\bftheta}$ and $\tilde\bX_{\hat\bftheta}$ requires $\order(nmp)$ time. If we require $L$ iterations going back and forth between estimating $\bfbeta$ and $\bftheta$, $L_{\theta}$ iterations to estimate $\bftheta$ given $\hat\bfbeta$, and $L_{\beta}$ iterations in the coordinate descent (including selecting tuning parameters using cross-validation) to estimate $\bfbeta$ given $\hat\bftheta$, the overall cost of our algorithm is $\order(n L(L_{\theta}m^3 + L_{\beta}p))$. We utilize a tolerance \texttt{tol} $ = 10^{-6}$ for the stopping criterion; a less stringent tolerance may be used, which will likely result in a smaller number of iterations $L$ and a less accurate approximation of $\bfbeta$ and $\bftheta$.


\section{Simulation study\label{sec:simulation}}

\subsection{Simulation scenarios\label{sec:scenarios}}

We sampled $2{,}000$ spatiotemporal coordinates from the possible combinations of 276 unique days and 50 unique spatial locations randomly distributed across the United States. The unique days correspond to the set of dates with complete geographic covariate datasets.  The 2,000 coordinates are randomly divided in half for a training set of size $n=1{,}000$, and a test set of size $n_P = 1{,}000$ that was never used in any model development. 

For the spatiotemporal regression errors $\epsilon_i$ in \eqref{eq:regressioni}, we specified a baseline (minimum; maximum) scenario of the model, with spatial range parameter $\gamma_s$ = 1,000 (200; 3,000) km, temporal range $\theta_t =$ 30 (7; 365) days, total variance (i.e., sill) $\sigma^{2}_{\text{total}} = \sigma^2 + \tau^2 = s^{2}_{\text{trend}}$ (0.5$\times s^{2}_{\text{trend}}$; 5$\times s^{2}_{\text{trend}}$) , and nugget-to-sill ratio $\tau^2/ \sigma^{2}_{\text{total}} = 0.25$ (0.01; 0.99), where $s^{2}_{\text{trend}}$ is the sample variance of the entries of the regression term $\bX\bfbeta$, evaluated at the true value of $\bfbeta$ (see below). The nugget-to-sill ratio is the ratio of the noise to total variance.
We considered a large number of simulation scenarios in which we varied, in turn, each of these variables, while holding the other variables fixed at their baseline levels. (Results for additional scenarios in which the variables varied jointly, including a spatial range of 30 km, are shown in Section \ref{Sec:Joint-Varying}.)

For the simulated spatiotemporal coordinates, we created spatiotemporal covariate matrices $\bX$ and $\bX_P$ based on the methods described for the NO$_2$ data in Section \ref{sec:data}, in order to obtain a realistic simulation setting. To provide a unique set of covariates from the NO$_2$ analysis, we included and removed variables as follows. Ozone TROPOMI satellite data (air mass factor (AMF), total column, and slant) were included with 1, 10, and 100 km buffers. Randomly generated point sources with isotropic exponentially decaying contribution with decay ranges of 1, 10, and 100 km \citep{Messier2012} and randomly generated spatiotemporal random fields were included in lieu of the NEI and road covariates from the NO$_2$ dataset. The final candidate set for the simulation included 123 potential covariates. The true trend coefficients $\bfbeta$ were assumed to have 8 non-zero coefficients: NO$_2$ Slant 1 km, cloud fraction 10 km, ozone AMF 100 km, precipitation 10 km, NDVI 100 km, developed high intensity 1 km, point sources with 100 km decay range, and a smoothly varying spatiotemporal random field; the corresponding true coefficient values were 5, 5, 3, -3, -5, 10, 3, 5, respectively. The true covariates exhibited low to moderate correlation ($|\rho|\leq 0.41$) with the other true covariates, and low to extremely high correlation ($|\rho| \leq 0.99$) with the extraneous covariates.


\subsection{Approaches under comparison\label{sec:Approaches}}

We compared our proposed method to several popular land-use regression approaches:

\begin{description}
  \item[LUR-iid:] An iid land-use regression model, which can be viewed as a special case of \eqref{eq:regressioni} with $\eta(\cdot) \equiv 0$. Point predictions are then simply given by $\bX_P\hat\bfbeta$.
  \item[LURK-Local:] Land-use regression Kriging with a local neighborhood, which is the current state-of-the-art approach in land-use regression \citep{DeHoogh2018}. LURK-local consists of the following steps:
    \begin{enumerate}[itemsep=0pt,topsep=0pt]
    \item Estimate $\hat\bfbeta$ as in LUR-iid, and compute residuals $\bfepsilon_{\hat\bfbeta}$.
    \item Estimate $\bftheta$ as an average of estimates based on $k=10$ samples of size $l= (m \frac{n}{k})^{1/3}$ from $\bfepsilon_{\hat\bfbeta}$ 
    \item Carry out local kriging at each prediction coordinate using the $m$ nearest (in terms of \eqref{eq:dist}) space-time neighbors among $\bfepsilon_{\hat\bfbeta}$.
    \end{enumerate}
\item[LURK-Vecchia:]  Our proposed methodology, summarized in Algorithm \ref{alg:lurkv}.
  \item[LURK-Full:] The full Kriging and SCAD penalized method based on \eqref{eq:lik}. This is equivalent to the proposed LURK-Vecchia approach with $m = n -1$.
\item[Local-Kriging:] Does not use geographic covariates. Similar to LURK-local, we estimate $\bftheta$ as an average of estimates based on $k=10$ samples of size $l= (m \frac{n}{k})^{1/3}$ from $\bz$, and then make predictions using the $m$ nearest spatiotemporal observations.
\end{description}
LURK-Full can be considered the most accurate approach, but it is computationally infeasible for large $n$ (in the tens of thousands or more). For the other spatiotemporal approaches, we ensure similar computational complexity by using the same $m=25$.

\subsection{Prediction scores\label{sec:scores}}

For the prediction at unobserved coordinates, we considered three proper scoring rules \citep[e.g.,][]{Gneiting2014} that all compare the true simulated test data $\by_P^\star$ to the predictive distribution $\adens( \by_P | \bz ) = \normal(\by_P|\bfmu_P,\bfSigma_P)$ (see Section \ref{sec:pred}) as approximated by each method. The mean squared error (MSE) is given by $(1/n_P) \sum_{i=1}^{n_P} (\by_{P,i}^\star -\bfmu_{P,i})^2$, the log-score is given by $-(1/n_P) \sum_{i=1}^{n_P} \normal(\by_{P,i}^\star|\bfmu_{P,i},\bfSigma_{P,ii})$, and the continuous ranked probability score is given by $(1/n_P) \sum_{i=1}^{n_P} \int (F_i(x) - 1\{\by_{P,i}^\star \leq x \})^2 dx $, where $F_i$ is the cumulative distribution function of $\normal(\bfmu_{P,i},\bfSigma_{P,ii})$.
Each score is averaged over 20 simulations.

\subsection{Simulation results}
\label{Sec:Sim-Results}
\subsubsection{Out-of-sample prediction\label{sec:Sim-Pred}}

\begin{figure}
\centering\includegraphics[width =1\linewidth]{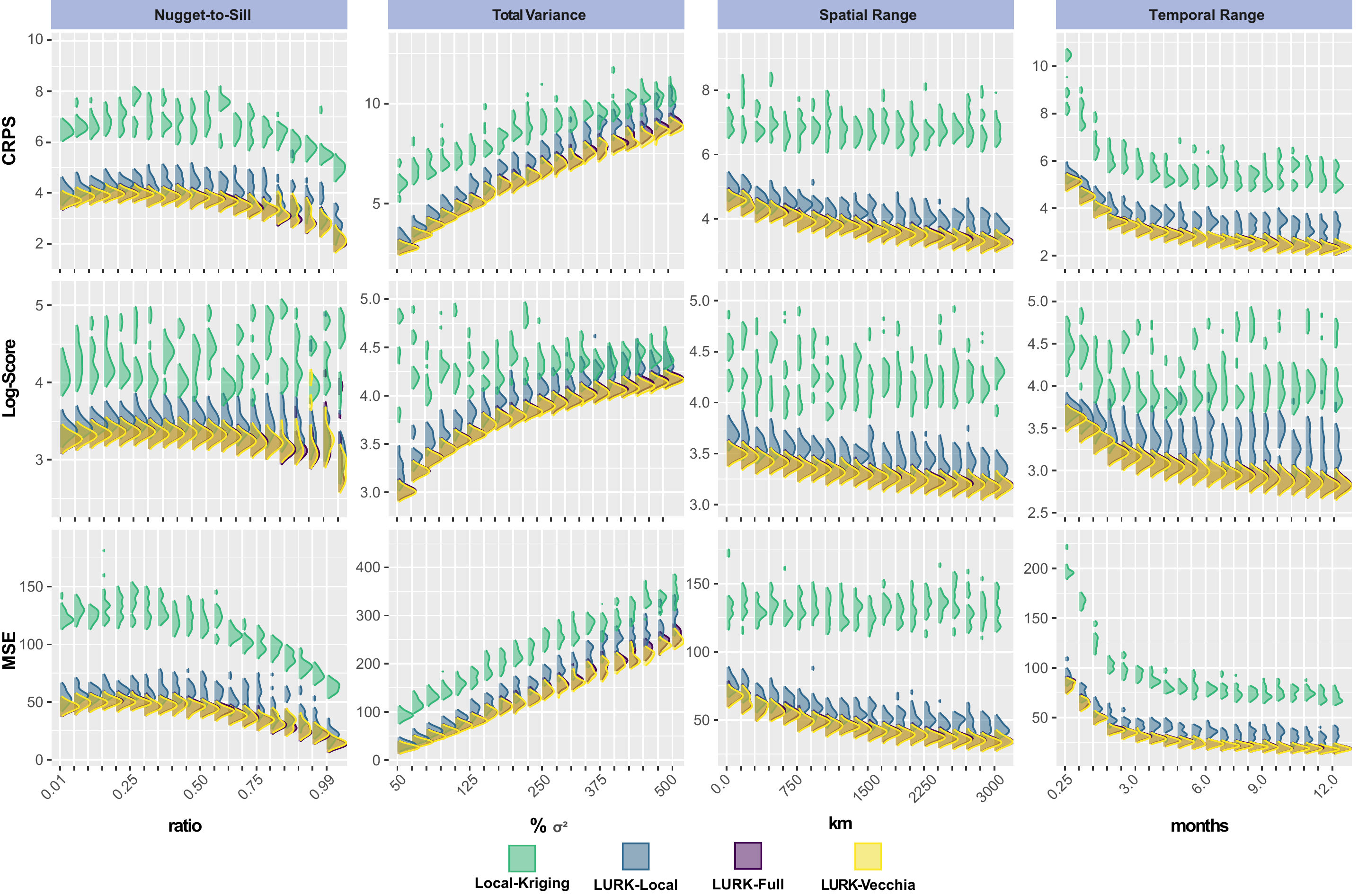}
  \caption{Ridgeline density plots of prediction scores (the lower the better) for different simulation scenarios (Section \ref{sec:scenarios}), in which the spatiotemporal parameters are singly varied with others held constant at a baseline level. 
  The LURK-Vecchia results were very close to and hence largely cover the results for LURK-Full.
  Note that each scenario has its own y-axis scale.}
\label{fig:Sim-ValStats-Curve}
\end{figure}

Figure \ref{fig:Sim-ValStats-Curve} shows ridgeline density plots of the prediction scores in Section \ref{sec:scores} for the methods in Section \ref{sec:Approaches} (except LUR-iid, which was not competitive) for the different simulation scenarios described in Section \ref{sec:scenarios}. The vertically oriented densities are generated as trimmed (inner 98 percent) density functions using the geom\_density\_ridges function in the ggplot2 and ggridges packages of R. The results for our LURK-Vecchia approach were nearly identical to those for LURK-Full. Across all scenarios, the average LURK-Vecchia scores were consistently between 5 and 60 percent better than those for LURK-Local. Local-Kriging was much worse.
(Plots showing percent differences are shown in Section \ref{Sec:Joint-Varying}.)

\subsubsection{Model selection\label{sec:Sim-Beta}}

\begin{figure}
\centering\includegraphics[width =1\linewidth]{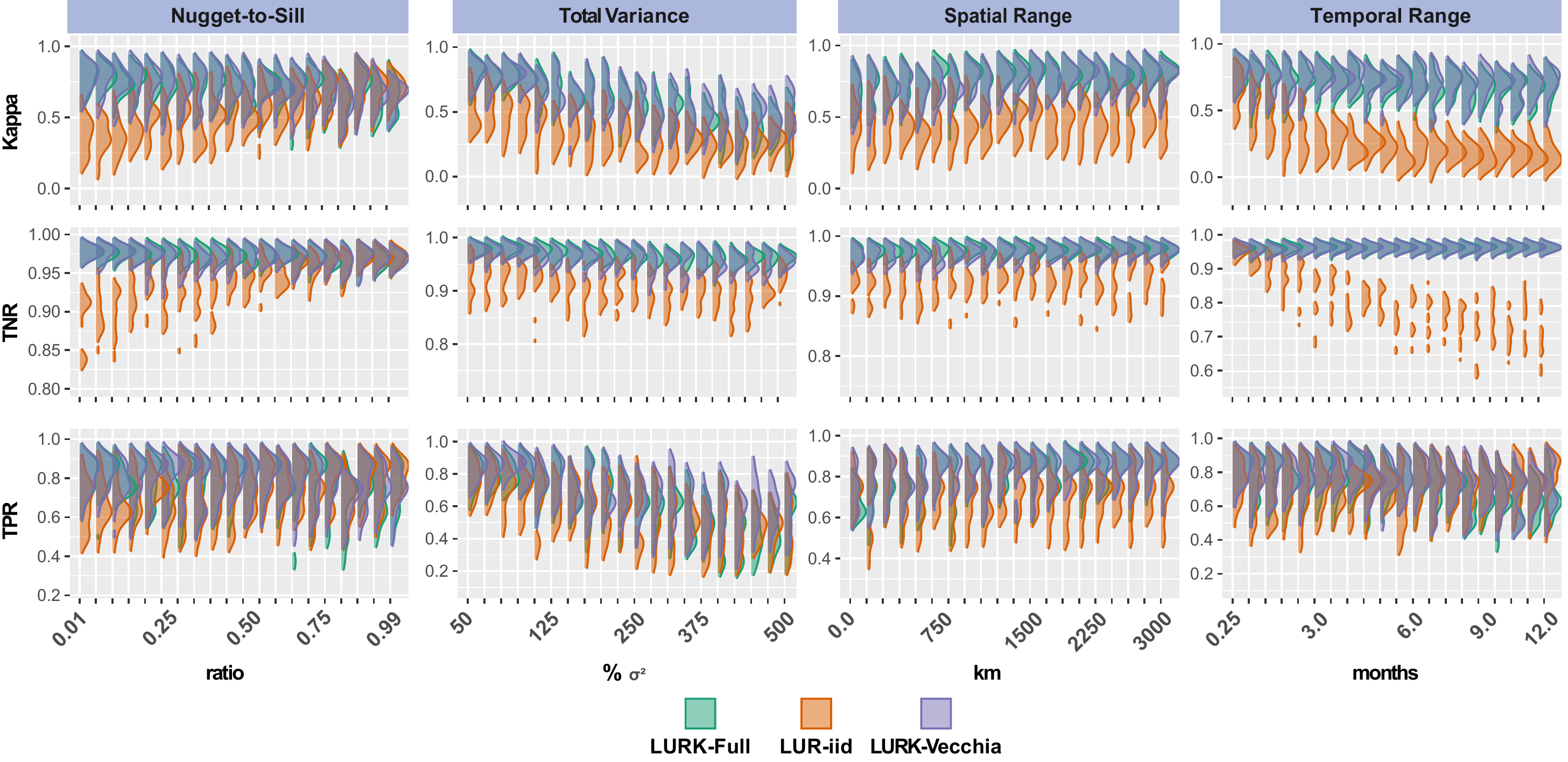}
  \caption{Ridgeline density plot comparison of model-selection statistics for simulation scenarios (Section \ref{sec:scenarios}), in which the spatiotemporal parameters are singly varied with others held constant at a baseline level.}
\label{fig:Sim-Betas-Curve}
\end{figure}

In terms of model selection, we considered the true negative rate (TNR), true positive rate (TPR), and Cohen's Kappa \citep{banerjee1999beyond}, $\kappa = {(p_o - p_e)}/{(1 - p_e)}$, where $p_o$ is the observed agreement of coefficient selections and $p_e$ is the expected agreement based on random chance.
Of the methods in Section \ref{sec:Approaches}, we omitted LURK-Local (because its model selection is identical to LUR-iid) and Local-Kriging (because it does not perform model selection).
Figure \ref{fig:Sim-Betas-Curve} shows the model-selection statistics for the scenarios in Section \ref{sec:scenarios}. Similar to the prediction scores, the distributions of the LURK-Vecchia and LURK-Full were very similar, indicating that the LURK-Vecchia approach approximated the full model well in terms of model selection. 
Compared to LUR-iid, LURK-Vecchia had 5 to 80 percent higher average TPR. The difference in terms of TNR and Kappa was even greater, with LUR-iid selecting a large number of erroneous non-zero coefficients (Figure \ref{fig:Sim-Betas-Nonzero-Coef}). The results were only comparable for scenarios with negligible spatiotemporal dependence (i.e., high nugget-to-sill ratio, or small ranges). 
An additional plot showing the percent differences in Kappa is shown in Section \ref{Sec:Joint-Varying}.

\section{Ground-level NO$_2$ analysis\label{sec:application}}
 
We now return to the daily, US-wide, ground-level NO$_2$ data described in Section \ref{sec:application}.  The minimum, maximum, mean (standard deviation), and median (interquartile range) observed concentrations were 0.004, 62.9, 8.5 (7.4), and 6.4 (9.0) parts-per-billion (ppb), respectively. Because NO$_2$ is positive and right-skewed, NO$_2$ was natural-log-transformed prior to the analyses.

\subsection{Comparison using cross-validation\label{s:cv}}

We compared the predictive accuracy using 10-fold cross-validation for the methods described in Section \ref{sec:Approaches}; LURK-Full was omitted because it is intractable for the large sample size. We used the same predictive scores as detailed in Section \ref{sec:scores}, except that we considered $\adens( \bz_P | \bz ) = \normal(\bz_P|\bfmu_P,\bfSigma_P+ \tau^2 \bI_{n_P})$, because the error-free $\by_P$ was unknown.

The results are shown in Table \ref{tab:NO2_CV}. LURK-Vecchia outperformed all other methods in terms of all three scores, resulting in a roughly ten percent decrease in MSE and log-score compared to the next best approach, LURK-Local. Further, LURK-Vecchia resulted in a 20 percent or greater decrease in all of the scores compared to Local-Kriging and LUR-iid. 

\begin{table}[tbp]
\small
\centering
\begin{tabular}{l|rrr}
  Method & MSE (ppb$^2$) & CRPS & Log-Score \\\hline
   Local-Kriging & 0.30 & 0.30 & 0.83 \\
  LUR-iid & 0.42 & 0.38 & 3.90 \\
 LURK-Local & 0.22 & 0.25 & 0.68 \\
 LURK-Vecchia & 0.20 & 0.24 & 0.61 \\
\end{tabular}
\caption{Predictive scores for NO$_2$ 10-fold cross-validation}
\label{tab:NO2_CV}
\end{table}


In terms of model selection, the mean (standard deviation) number of non-zero coefficients was 71 (1.1) and 24 (1.7) for LUR-iid and LURK-Vecchia, respectively. The LUR-iid models were severely affected by multicollinearity; the median (mean) variance inflation factor (VIF) for all 10 cross-validation models was 2.3 (6.7) and 5.7 (18.6) for LURK-Vecchia and LUR-iid, respectively. LURK-Local uses the same model-selection procedure as LUR-iid. Other iid model-selection approaches are likely to result in similarly large numbers of covariates \citep[e.g.,][]{kerckhoffs2019}.

Thus, by appropriately accounting for spatiotemporal dependence, LURK-Vecchia resulted in more accurate, sparser, and more interpretable models than LUR-iid and LURK-Local.

\subsection{Prediction maps\label{sec:Predictionmaps}}

Having shown using cross-validation that our proposed LURK-Vecchia approach can outperform the competing methods, we fitted LURK-Vecchia to the entire dataset. The covariance parameters $\bftheta = (\sigma^2,\gamma_s,\gamma_t,\tau^2)$ were estimated as (2.2  ppb$^2$, 1.4 km, 0.63 yr, 0.15 ppb$^2$), and the trend coefficients are given in Table \ref{tab:NO2_CV} and discussed in Section \ref{sec:no2vars}. The entire estimation algorithm required $L=4$ iterations and took approximately 86 minutes on a machine with 16GB RAM and an Intel(R) i7-8665U processor (4 cores, 1.90GHz). 

\begin{figure}
\centering\includegraphics[width =1\linewidth]{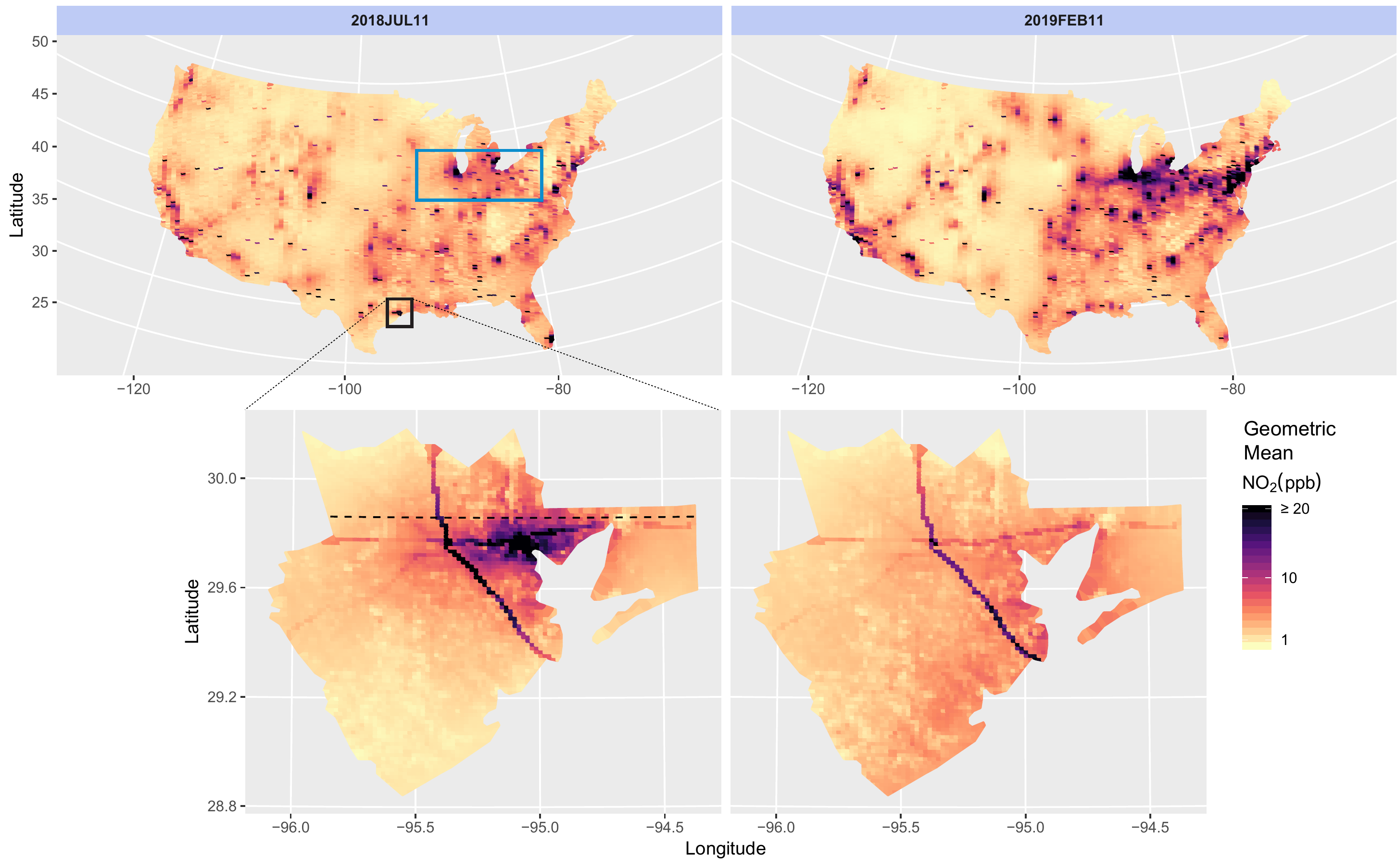}
  \caption{LURK-Vecchia prediction geometric mean (exponentiated log-scale predictions) across the conterminous United States (top row) and a five-county subset around Houston, Texas, for July 11, 2018, and February 11, 2019. The horizontal transect in the Texas subfigure is used in Figure \ref{fig:Transect} to show spatiotemporal patterns in more detail.}
\label{fig:VecchiaPred}
\end{figure}

Figure \ref{fig:VecchiaPred} shows the prediction geometric mean, $\exp(\bfmu_P)$, for two distinct days, for the entire US domain and for a more detailed 5-county area surrounding Houston, Texas.
(Corresponding prediction uncertainties are shown in Figure \ref{fig:VecchiaSD}.) For the US domain, predictions were produced on a 200 by 100 grid (10--20 km resolution) across the conterminous United States. For evaluating fine-scale prediction patterns, a 1--2 km grid was produced in the Houston, TX five-county area. Distinct spatiotemporal patterns emerged. Cities and developed areas, such as roadways, showed elevated NO$_2$ concentrations, as expected for a traffic-related pollutant. However, there was temporal variability in the spatial patterns around the cities and roads. Comparing the the upper, mid-west cities, such as Chicago and Cleveland (blue box, Figure \ref{fig:VecchiaPred}), predicted NO$_2$ was lower on July 11, 2018, than on February 11, 2019. In contrast, in the Houston area sub-figure, predicted NO$_2$ was higher on July 11, 2018, than on February 11, 2019. Visually inspecting the predictors in the final model can reveal the primary drivers of the spatiotemporal variability, which is easy in linear models with interpretable covariates. Complex machine learning models and dimension-reduction techniques do not allow for such intuitive visual comparisons. Figure \ref{fig:Pred-Covariate-Comparison} shows the predictions and select covariates in the Houston area on July 11, 2018 and February 11, 2019. Visual inspection and correlation of covariate-only predictions with the final predictions show that the TROPOMI NO$_2$ data are driving the patterns observed on July 11, 2018. Contrarily, on Feb 11, 2019 other factors such as the specific humidity display similar general spatial patterns and have high correlation with the final predictions.

\begin{figure}[!htbp]
\centering\includegraphics[width =.8\linewidth]{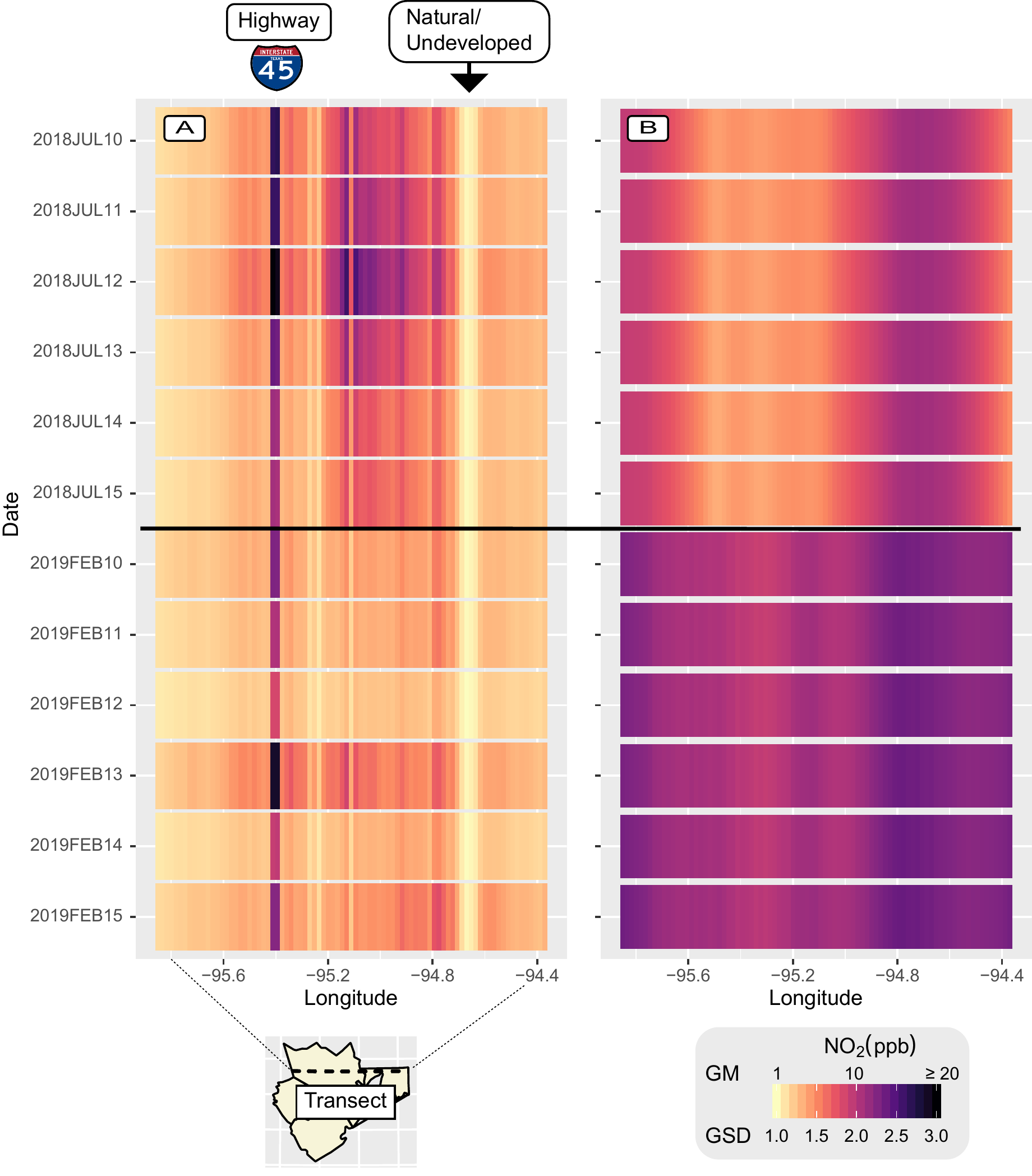}
  \caption{For LURK-Vecchia predictions, geometric mean (A) and geometric standard deviation (B) for two sets of dates (y-axis) and along a transect (x-axis) shown in the Houston panel of Figure \ref{fig:VecchiaPred}}
\label{fig:Transect}
\end{figure}

Figure \ref{fig:Transect} shows spatiotemporal predictions in more detail for a transect through the Houston panel of Figure \ref{fig:VecchiaPred}. Moving along the transect, we see a general spatial pattern with modulations in time. For instance, the highway consistently had the highest observed concentrations, but the magnitude of the maximum fluctuated daily, driven by time-varying covariates such as the TROPOMI and meteorological variables (see Section \ref{sec:no2vars}). July 12, 2018 had consistently higher concentrations than other days across most of the transect locations, including the highway. We also showed prediction uncertainties in terms of geometric standard deviations (SDs), $\exp\big(\diag(\bfSigma_P)^{1/2}\big)$. The SD varied over longer time periods than the mean, as evidenced by minor differences within the 2018 and 2019 ranges, but considerable differences between them. 

Our predictions will be useful for epidemiological and risk-assessment studies seeking daily, national-scale predictions. For example, \citet{mills2015quantitative} provide meta-analysis results for the impacts of 24-hour NO$_2$ exposure on all-age-group, all-cause mortality, cardiovascular mortality, respiratory mortality, cardiovascular hospital admissions, and respiratory hospital admissions. Our daily NO$_2$ exposure predictions, combined with population information and with the \citet{mills2015quantitative} relative-risk estimates, may be used to develop NO$_2$ acute-health impact assessments, such as an attributable-fraction of mortality. Please contact the authors to request predictions at the desired spatiotemporal coordinates.

\subsection{Interpretation of selected covariates\label{sec:no2vars}}

\begin{table}[!htbp]
\small
\begin{tabular}{ll|r|r|r}
 Variable Category & Variable Name & res.\ (km) & $\hat{\beta}$ & $(e^{\hat{\beta}}-1)$ $\times$  100  \\
 \hline
 \hline
 Intercept   & - & - & 0.002 & -\\
 \hline
TROPOMI (Sec \ref{sec:TROPOMI})  &  NO$_2$ Slant & 1 & 0.086 &8.9\\ 
          &  NO$_2$ Slant & 100 & -0.002 &-0.21 \\
          & NO$_2$ Tropospheric & 10 & 0.036 & 3.6 \\
          & NO$_2$ Tropospheric & 100 & 0.044 & 4.5\\
          & Tropo Pressure & 1 & -0.003 & -0.31\\
          & Tropo Pressure & 100 & -0.0004 & -0.04 \\
          & AAI & 100 & -0.025 & -2.6\\
         & cloud fraction & 1 & 0.057 & 5.9\\
 \hline
Meteorology (Sec \ref{sec:Meteorology})  &  rhmax & 1 & 0.12 & 13.1\\
          &  rhmax & 100 & -0.090 & -8.6 \\
          & sph & 100 & -0.27 & -23.6\\
          & tmax & 100 & 0.44 & 55.5\\
          & vs & 100 & -0.21 & -19.2 \\
 \hline
Vegetation  (Sec \ref{sec:veg})  &  NDVI & 1 & -0.052 & -5.1\\
 \hline
Population \& Roads (Sec \ref{sec:Population})  & Travel Friction  & 10 & -0.038 & -3.7\\
          &  Total Road & 1 & 0.18 & 20.3\\
 \hline
Land cover  (Sec \ref{sec:landcover})  &  Water & 1 & 0.0078 & 0.79\\
          &  Mixed Forest & 1 & -0.21& -18.3\\
          & Mixed Forest & 10 & 0.017  & 1.7\\
          & Shrub & 1 & -0.035 & -3.5\\
          & Herbaceous & 1 & -0.081 & -7.8\\
          & Dev Open & 10 & 0.0031 & 0.31\\
          & Dev Low & 10 & 0.056 & 5.7\\
          & Elevation & 1 & -0.28 & -24.6\\
 \hline
Emissions  (Sec \ref{sec:NEI})  & NEI  & 1 & 0.0092 & 0.92
\end{tabular}
\caption{Selected covariates for NO$_2$ data: rhmax = maximum relative humidity, vs = wind velocity, sph = specific humidity, tmax = maximum temperature. For every 1-standard-deviation increase in a covariate, we expect an estimated $(e^{\hat\beta}-1)\times 100$ percent increase in NO$_2$.}
\label{tab:NO2_Beta}
\end{table}

Table \ref{tab:NO2_Beta} shows the 25 variables selected by our LURK-Vecchia procedure, along with their estimated coefficients. We now discuss interpretations and context for each selected variable, grouped by variable category (see Section \ref{sec:no2vars}): 
\begin{itemize}
    \item \textbf{TROPOMI}. Similar to many LUR studies \citep{Novotny2011,Young2016, Larkin2017, DeHoogh2018}, we found satellite observations of NO$_2$ selected to the LURK-Vecchia model. The two covariates for NO$_2$ Slant combine for a net positive effect on ground-level NO$_2$ while providing moderation between local and regional scale effects from the 1 and 100 km circular buffer hyperparameters, respectively. Similarly, the tropospheric NO$_2$ variables have two variables at different spatial scales that contribute a net positive effect to ground-level NO$_2$. 
    
    The TROPOMI variables for tropopause pressure at 1 and 100 km were selected with small negative coefficients, indicating areas of reduced ground-level NO$_2$. These variables, which have a relatively small impact on the final prediction concentrations, may represent a minor changes in the mixing volume for pollutant molecules. 
    
    The dynamic relationship between NO$_2$ and aerosols is complex and not completely understood. For instance, \citet{grundstrom2015} observe weak to moderate correlations between total NO$_x$ (NO + NO$_2$) and particle number concentrations (PNC) depending on meteorological conditions such as wind velocity. \citet{apte2019potential} found PNC to have a consistent diurnal pattern of midday new particle formation that is poorly approximated with NO$_x$. Without a priori expectations of the AAI coefficient, we observe a 2.6 percent decrease in NO$_2$ concentrations for every one SD increase in AAI. We find a positive coefficient for cloud fraction, which is likely due to the protective effect of clouds on incoming solar radiation. 
    
     \item \textbf{Meteorology}. Two relative humidity and one specific humidity variable contribute a net negative effect on NO$_2$ concentrations. Similarly to AAI, we expect that water vapor and aerosols to impede solar radiation and breakdown of NO$_2$ to NO. 
     
     We observe a 55.5 percent increase in NO$_2$ concentrations for every 1 SD increase in the maximum daily temperature. Hot days are associated with increased solar radiation and O$_3$ formation.The significant increase in NO$_2$ concentrations is likely capturing O$_3$ mediated conversion of NO ($\ce{NO ->[O_3] NO_2}$) \citep{seinfeld_pandis}. 
     
     We observe a 19.2 percent decrease in NO$_2$ concentration with a 1 SD increase in wind velocity, which is expected as this increases transport of NO$_2$ and its precursors from the given location. 
     
     \item \textbf{Vegetation}. For every 1 SD increase in NDVI, we observe a 5.1 percent decrease in NO$_2$ concentrations. NDVI represents vegetative greenness, thus this is consistent with the lack of NO$_2$ or NO$_x$ sources. 
     
     \item \textbf{Land cover}. Open water has a small, positive impact on NO$_2$, which is likely due to the concentration of cities and sources near water sources and coastlines or as a proxy variable for ports. Mixed forest (the net sum of short and medium ranges), shrub-land, and herbaceous wetlands have negative contributions to NO$_2$ predictions, which is expected due to the lack of sources. Developed open and low have positive coefficients, while developed low is larger as it represents an increased anthropogenic presence. 
     
     For every 1 SD increase in elevation in 1 km buffer, there is a 24.6 percent decrease in NO$_2$ concentrations, which is consistent with other LUR models of NO$_2$ \citep{DeHoogh2018} and can be due to a combination of atmospheric mixing, fewer sources, decreased average temperature, and increased wind velocity at higher elevations. 
     
     \item \textbf{Population \& Roads}. We find travel friction within a 10 km buffer and total road length within a 1 km buffer to result in a 3.7 decrease and 20.3 increase of NO$_2$ concentrations for every 1 SD increase, respectively. Travel friction is the average travel time, and total road length is a good approximation of vehicle sources, and so they are expected to decrease and increase traffic-related pollutants, respectively. 
     
     \item \textbf{Emissions}. The NEI variable with a 1 km decay range was selected with a small, positive coefficient. Clearly, we expect a covariate representing source emissions of the dependent variable to be contribute positively.
     
\end{itemize}

\section{Conclusions\label{sec:conclusions}}


We analyzed daily ground-level NO$_2$ concentrations across the United States, using a novel penalized land-use regression approach with spatiotemporally correlated errors that is also scalable to large datasets via a sparse general Vecchia approximation. Our methodological advances can be used in future human health exposure and risk assessment to improve model selection and prediction characteristics. Key results from the NO$_2$ analysis include: the development of daily NO$_2$ concentration predictions that can be used for epidemiological analyses of acute health effects such as asthma and increased hospitalizations; the potential to develop annual average concentrations that propagate uncertainty from daily predictions as opposed to those based on direct annual averages; the elucidation of spatiotemporal patterns of NO$_2$ concentrations across the United States, including significant variations between cities and intra-urban variation; and the resolving of a parsimonious group of geographic covariates describing the spatiotemporal distribution of daily NO$_2$ concentrations, including satellite imagery, meteorological data, land cover, population distributions, road networks, and point source emissions.  


Our methods also offer a scalable way to analyze other large spatiotemporal datasets in environmental and human health risk assessment. For example, in the air-quality research community, mobile monitoring of air pollutants is leading to high-resolution datasets with millions of observations, including campaigns in Zurich, Switzerland \citep{li2012sensing}, Boston, MA \citep{padro2012mobile}, Oakland, CA \citep{apte2017high,Guan2020}, Houston, TX \citep{miller2020characterizing}, and the Netherlands \citep{kerckhoffs2019}. 

Our methods could also be extended to non-Gaussian data \citep{Zilber2019} or on-line spatiotemporal filtering \citep{Jurek2018} using extensions or variations of the general-Vecchia framework.

As mentioned in Section \ref{sec:lur}, the SCAD penalty used for model selection in Line \ref{l:scad} of Algorithm \ref{alg:lurkv} could be replaced by other penalties, such as LASSO \citep{Tibshirani1996}, elastic net \citep{Zou2005}, or relaxed LASSO \citep{hastie2017extended}, which may result in improvements in prediction accuracy or model selection.
While accurate uncertainty quantification and significance assessment is difficult in the context of penalized regression, a potential extension of our approach would be to combine it with existing methods proposed for this purpose \citep[e.g.,][]{Meinshausen2009,chatterjee2011,Xie2019}. This would likely come at an increased computational cost, but it would also allow for the inclusion of covariate uncertainty in predictions. Lastly, another possible avenue is to adjust for spatial confounding as proposed in \citet[][]{Hughes2013}.

\footnotesize
\appendix
\section*{Acknowledgments}

Messier's research was partially conducted while at Oregon State University, Department of Environmental and Molecular Toxicology, and supported by NIEHS K99 ES029523. Messier is currently supported by NIH institutes NIEHS/NTP and NIMHD as an intramural investigator.
Katzfuss' research was partially supported by National Science Foundation (NSF) Grants DMS--1654083 and DMS--1953005. Simulations were run on computing resources at the Oregon State University Center for Genome Research and Biocomputing.
The authors would like to thank Shahzad Gani, Jianhua Huang, Irina Gaynanova, Anirban Bhattacharya, and Joe Guinness for helpful comments and suggestions.

\section{Review of general Vecchia \label{app:vecchia}}

We now provide some further details of the general Vecchia approximation \citep{Katzfuss2017a,Katzfuss2018} that we extended and briefly reviewed in Section \ref{sec:methodology}. Because model \eqref{eq:regressioni} implies conditional independence in \eqref{eq:vecchiaapprox} between $z_i$ and all other variables in $\bu$ given $y_i$, we assume that $z_i$ always conditions on only $y_i$. Hence, we can write the approximation \eqref{eq:vecchiaapprox} as
\[
\adens(\bu)  = \prod_{i=1}^{n} p(z_i | y_{i} ) \, p(y_i | \by_{q_y(i)}, \bz_{q_z(i)}),
\]
where $q(i) = q_y(i) \cup q_z(i)$ with $q(i) \subset (1,\ldots,i-1)$ is the conditioning index vector of size $|q(i)|\leq m$, and we assume $q_y(i) \cap q_z(i) = \emptyset$.
The ordering of the variables and the choice of conditioning sets can have a strong effect on the approximation accuracy and computational speed.

The ordering of the spatiotemporal coordinates $(\bs_1,t_1),\ldots,(\bs_n,t_n)$ implies an ordering of the variables in $\bu$. We assume here that the coordinates are ordered and numbered according to a maximum-minimum distance ordering \citep{Guinness2016a}, which sequentially picks each coordinate in the ordering to maximize the minimum distance to previous coordinate in the ordering. The conditioning index vectors $q(i)$ are chosen here as the indices of the nearest $m$ coordinates previous to $i$ in this ordering. To determine the ordering and the conditioning sets, we use the scaled spatiotemporal distance \eqref{eq:dist} as our measure of distance \citep[cf.][]{Datta2016a}. However, this measure of distance depends on the unknown parameters $\bftheta$ (specifically, on $\gamma_s$ and $\gamma_t$), and so we update the ordering and conditioning at each iteration (in Line \ref{l:ocupdate}) of Algorithm \ref{alg:lurkv} based on the current estimate of $\bftheta$.

Different strategies for splitting $q(i)$ into $q_y(i)$ and $q_z(i)$ can also result in vastly different approximation accuracies. In general, conditioning on $y_j$ is often more accurate but also potentially more computationally expensive than conditioning on $z_j$. \citet{Katzfuss2017a} proposed a fast and accurate sparse general Vecchia (SGV) approach that chooses $q_y(i) \subset q(i)$ such that $j<k$ can only both be in $q_y(i)$ if $j \in q_y(k)$, with the remaining conditioning indices in $q(i)$ assigned to $q_t(i) = q(i) \setminus q_y(i)$. Specifically, for $i=1,\ldots,n$, SGV finds $p(i) = \argmax_{j \in q(i)} |q_y(j) \cap q(i)|$ and $k_i=\argmin_{\ell \in p(i)} \|\bs_i - \bs_\ell\|$, and then sets $q_y(i) = (k_i) \cup (q_y(k_i) \cap q(i))$. We use SGV in all our numerical examples.

The restriction of conditioning only on previous variables in the ordering, $q_y(i) \subset (1,\ldots,i-1)$, ensures that the implied joint distribution is multivariate normal as indicated in \eqref{eq:vecchiaapprox}.
To compute the sparse upper-triangular matrix $\bU$, let $g(i)$ denote the vector of indices of the elements in $\bu$ on which $u_i$ conditions (e.g., if $u_i = z_k$ then $g(i) = (i-1)$). Also define $K(y_i,y_j) = K(z_i,y_j) = C\big((\bs_i,t_i),(\bs_j,t_j)\big)$ and $K(z_i,z_j) = C\big((\bs_i,t_i),(\bs_j,t_j)\big) + \mathbbm{1}_{i=j} \tau^2_i$. Then, the $(j,i)$th element of $\bU$ is
\begin{equation*}
\label{eq:U}
\bU_{ji} = \begin{cases} r_i^{-1/2}, & i=j,\\ -b_{i}^{(j)} r_i^{-1/2}, & j \in g(i), \\ 0, &\textnormal{otherwise}, \end{cases}
\end{equation*}
where $\bb_i'= K(u_i,\bu_{g(i)}) K(\bu_{g(i)},\bu_{g(i)})^{-1}$, $r_i = K(u_i,u_i) - \bb_i'K(\bu_{g(i)},u_i)$, and $b_i^{(j)}$ denotes the $\ell$th element of $\bb_i$ if $j$ is the $\ell$th element in $g(i)$ (i.e., $b_i^{(j)}$ is the element of $\bb_i$ corresponding to $u_j$).


For prediction of $\by_P$, we employ a spatio-temporal extension of the response-first ordering full-conditioning (RF-full) approach of \citet{Katzfuss2018}, which applies a general Vecchia approximation of the form \eqref{eq:vecchiaapprox} to $\tilde\bu=(\bz',\by_{\text{all}}')'$, where $\by_{\text{all}} = (\by',\by_P')' \equalscolon (y_1,\ldots,y_{n_{\text{all}}})'$. This results in the approximation
\[
\adens(\bz,\by_{\text{all}}) = \prod_{i=1}^n \dens(z_i) \times \prod_{i=1}^{n_{\text{all}}} \dens(y_i\vert \by_{q_y(i)}, \bz_{q_z(i)}),
\]
where $\by_{q_y(i)}$ and $\bz_{q_z(i)}$ are chosen as the $m$ variables closest in scaled distance \eqref{eq:dist} to $y_i$, among those that are previously ordered in $\tilde\bu$, where we condition on $y_j$ instead of $z_j$ whenever possible. Specifically, we set $q(i)$ to consist of the indices corresponding to the $m$ nearest spatiotemporal coordinates, including $i$ for $i \leq n $, and not including $i$ for $i >n$. Then, for any $j \in q(i)$, we let $y_i$ condition on $y_j$ if it is ordered previously in $\bu$, and condition on $z_j$ otherwise. More precisely, we set $q_y(i) = \{j \in q(i): j<i\}$ and $q_z(i) = \{j \in q(i): j\geq i\}$.
 

\section{Alternative expression of the objective function \label{app:objective}}

We now show that the objective function $\hat Q(\bfbeta,\bftheta)$ can be written as in \eqref{eq:qvecchia} as:
\[
\textstyle\hat Q(\bfbeta,\bftheta) = \|\tilde\bz_{\bftheta} - \tilde\bX_{\bftheta}\bfbeta\|_2^2 + \lambda\, \pen(\bfbeta) - 2 \sum_{i} \log \big((\bU_{\bftheta})_{ii}\big) + 2\sum_{i} \log \big((\bV_{\bftheta})_{ii}\big).
\]
As in \citet[][proof of Prop.~2]{Katzfuss2017a}, note that, for any value of $\by$,
$
\adens(\bz) = \adens(\bu)/\adens(\by|\bz),
$
where $\adens(\bu)$ is given in \eqref{eq:vecchiaapprox}, and $\adens(\by|\bz) = \normal_n(\bfmu,\bW^{-1})$ with $\bfmu = \bX\bfbeta - \bW^{-1}\bA\bB'(\bz-\bX\bfbeta)$. Thus, setting $\by=\bfmu$, and denoting by $\bu_{\bfmu} = (\mu_1,z_1,\ldots,\mu_n,z_n)$ the resulting vector $\bu$, we obtain
\begin{align*}
-2 \log \adens(\bz) 
  & = -2 \log \normal_{2n}(\bu_{\bfmu}|(\bX \kronecker \mathbf{1}_2)\bfbeta,\hat\bfSigma) + 2 \log  \normal_n(\bfmu|\bfmu,\bW^{-1}) \\
  & = (\bu_{\bfmu}-(\bX \kronecker\mathbf{1}_2)\bfbeta)'\bU\bU'(\bu_{\bfmu}-(\bX \kronecker \mathbf{1}_2)\bfbeta) - \log |\bU\bU'| + \log |\bW| \\
  & =  \textstyle \|\bd\|^2_2 - 2 \sum_{i} \log \bU_{ii} + 2\sum_{i} \log \bV_{ii},
\end{align*}
where
\begin{align*}
\bd
  & = \bU'\bu_{\bfmu} - \bU'(\bX \kronecker \mathbf{1}_2)\bfbeta = \bB'\bz + \bA'\bfmu - (\bB'\bX + \bA'\bX)\bfbeta \\
  & = \bB'\bz - \bA'\bW^{-1}\bA\bB'\bz - (\bB'\bX + \bA'\bX - \bA'\bX + \bA'\bW^{-1}\bA\bB'\bX) \bfbeta \\
  & = \tilde\bz - \tilde\bX\bfbeta,
\end{align*}
where $\tilde\bz = \bB'\bz + \bA'\bW^{-1} \bA\bB'\bz$ and $\tilde\bX = \bB'\bX + \bA'\bW^{-1}\bA\bB'\bX$.

\footnotesize
\bibliographystyle{apalike}
\bibliography{mendeley,additionalrefs}

\begin{thebibliography}{}

\bibitem[Abatzoglou et~al., 2014]{abatzoglou2014}
Abatzoglou, J.~T., Rupp, D.~E., and Mote, P.~W. (2014).
\newblock {Seasonal climate variability and change in the Pacific Northwest of
  the United States}.
\newblock {\em Journal of Climate}, 27(5):2125--2142.

\bibitem[Alexeeff et~al., 2018]{Alexeeff2018}
Alexeeff, S.~E., Roy, A., Shan, J., Liu, X., Messier, K., Apte, J.~S., Portier,
  C., Sidney, S., and {Van Den Eeden}, S.~K. (2018).
\newblock {High-resolution mapping of traffic related air pollution with Google
  Street View cars and incidence of cardiovascular events within neighborhoods
  in Oakland, CA}.
\newblock {\em Environmental Health}, 17(1):1--13.

\bibitem[Apte et~al., 2019]{apte2019potential}
Apte, J., Gani, S., Chambliss, S., Messier, K., Lunden, M., et~al. (2019).
\newblock Potential underestimation of ultrafine particle exposure when using
  proxy pollutants: Lessons from long-term measurements at fixed sites and
  mobile monitoring.
\newblock {\em Environmental Epidemiology}, 3:13--14.

\bibitem[Apte et~al., 2017]{apte2017high}
Apte, J.~S., Messier, K.~P., Gani, S., Brauer, M., Kirchstetter, T.~W., Lunden,
  M.~M., Marshall, J.~D., Portier, C.~J., Vermeulen, R.~C., and Hamburg, S.~P.
  (2017).
\newblock High-resolution air pollution mapping with google street view cars:
  exploiting big data.
\newblock {\em Environmental Science \& Technology}, 51(12):6999--7008.

\bibitem[Banerjee et~al., 1999]{banerjee1999beyond}
Banerjee, M., Capozzoli, M., McSweeney, L., and Sinha, D. (1999).
\newblock Beyond kappa: A review of interrater agreement measures.
\newblock {\em Canadian journal of statistics}, 27(1):3--23.

\bibitem[Beckerman et~al., 2013]{Beckerman2013}
Beckerman, B.~S., Jerrett, M., Serre, M.~L., Martin, R.~V., Lee, S.-j.,
  Donkelaar, A.~V., Ross, Z., Su, J., and Burnett, R.~T. (2013).
\newblock {A hybrid approach to estimating national scale spatiotemporal
  variability of PM$_{2.5}$ in the contiguous United States}.
\newblock {\em Environmental Science {\&} Technology}, 47(13):7233--7241.

\bibitem[Boersma et~al., 2007]{boersma2007}
Boersma, K., Eskes, H., Veefkind, J., Brinksma, E., Van Der~A, R., Sneep, M.,
  Van Den~Oord, G., Levelt, P., Stammes, P., Gleason, J., et~al. (2007).
\newblock {Near-real time retrieval of tropospheric NO$_2$ from OMI}.
\newblock {\em Atmospheric Chemistry and Physics}, 7(8):2103--2118.

\bibitem[Breheny and Huang, 2011]{breheny2011}
Breheny, P. and Huang, J. (2011).
\newblock Coordinate descent algorithms for nonconvex penalized regression,
  with applications to biological feature selection.
\newblock {\em The Annals of Applied Statistics}, 5(1):232.

\bibitem[Briggs et~al., 1997]{Briggs1997}
Briggs, D.~J., Collins, S., Elliott, P., Fischer, P., Kingham, S., Lebret, E.,
  Pryl, K., {Van Reeuwijk}, H., Smallbone, K., and {Van Der Veen}, A. (1997).
\newblock {Mapping urban air pollution using GIS: a regression-based approach}.
\newblock {\em International Journal of Geographical Information Science},
  11(February 2015):699--718.

\bibitem[{Center for International Earth Science Information Network - CIESIN -
  Columbia University}, 2018]{GPWv4}
{Center for International Earth Science Information Network - CIESIN - Columbia
  University} (2018).
\newblock Gridded population of the world, version 4 (gpwv4): Population
  density, revision 11 [data set].

\bibitem[Chatterjee and Lahiri, 2011]{chatterjee2011}
Chatterjee, A. and Lahiri, S.~N. (2011).
\newblock Bootstrapping lasso estimators.
\newblock {\em Journal of the American Statistical Association},
  106(494):608--625.

\bibitem[Coulliette et~al., 2009]{Coulliette2009}
Coulliette, A.~D., Money, E.~S., Serre, M.~L., and Noble, R.~T. (2009).
\newblock {Space/time analysis of fecal pollution and rainfall in an eastern
  North Carolina estuary.}
\newblock {\em Environmental Science {\&} Technology}, 43(10):3728--35.

\bibitem[Datta et~al., 2016a]{Datta2016}
Datta, A., Banerjee, S., Finley, A.~O., and Gelfand, A.~E. (2016a).
\newblock {Hierarchical nearest-neighbor Gaussian process models for large
  geostatistical datasets}.
\newblock {\em Journal of the American Statistical Association},
  111(514):800--812.

\bibitem[Datta et~al., 2016b]{Datta2016a}
Datta, A., Banerjee, S., Finley, A.~O., Hamm, N. A.~S., and Schaap, M. (2016b).
\newblock {Non-separable dynamic nearest-neighbor Gaussian process models for
  large spatio-temporal data with an application to particulate matter
  analysis}.
\newblock {\em Annals of Applied Statistics}, 10(3):1286--1316.

\bibitem[de~Hoogh et~al., 2018]{DeHoogh2018}
de~Hoogh, K., Chen, J., Gulliver, J., Hoffmann, B., Hertel, O., Ketzel, M.,
  Bauwelinck, M., van Donkelaar, A., Hvidtfeldt, U.~A., Katsouyanni, K.,
  Klompmaker, J., Martin, R.~V., Samoli, E., Schwartz, P.~E., Stafoggia, M.,
  Bellander, T., Strak, M., Wolf, K., Vienneau, D., Brunekreef, B., and Hoek,
  G. (2018).
\newblock {Spatial PM$_{2.5}$, NO$_2$, O$_3$, and BC models for Western Europe
  – Evaluation of spatiotemporal stability}.
\newblock {\em Environment International}, 120(2):81--92.

\bibitem[de~Hoogh et~al., 2019]{de2019predicting}
de~Hoogh, K., Saucy, A., Shtein, A., Schwartz, J., West, E.~A., Strassmann, A.,
  Puhan, M., Roosli, M., Stafoggia, M., and Kloog, I. (2019).
\newblock Predicting fine-scale daily no2 for 2005--2016 incorporating omi
  satellite data across switzerland.
\newblock {\em Environmental science \& technology}, 53(17):10279--10287.

\bibitem[Efron et~al., 2004]{Efron2004}
Efron, B., Hastie, T., Johnstone, I., and Tibshirani, R. (2004).
\newblock {Least angle regression}.
\newblock {\em Annals of Statistics}, 32(2):407--499.

\bibitem[Fan and Li, 2001]{FanLi2001}
Fan, J. and Li, R. (2001).
\newblock {Variable selection via nonconcave penalized likelihood and its
  oracle properties}.
\newblock {\em Journal of the American Statistical Association},
  96(456):1348--1360.

\bibitem[Finley et~al., 2009]{Finley2009}
Finley, A.~O., Sang, H., Banerjee, S., and Gelfand, A.~E. (2009).
\newblock {Improving the performance of predictive process modeling for large
  datasets}.
\newblock {\em Computational Statistics {\&} Data Analysis}, 53(8):2873--2884.

\bibitem[Friedman et~al., 2010]{Friedman2010}
Friedman, J., Hastie, T., and Tibshirani, R. (2010).
\newblock Regularization paths for generalized linear models via coordinate
  descent.
\newblock {\em Journal of Statistical Software}, 33(1):1--22.

\bibitem[Gauderman et~al., 2005]{gauderman2005childhood}
Gauderman, W.~J., Avol, E., Lurmann, F., Kuenzli, N., Gilliland, F., Peters,
  J., and McConnell, R. (2005).
\newblock Childhood asthma and exposure to traffic and nitrogen dioxide.
\newblock {\em Epidemiology}, pages 737--743.

\bibitem[Gneiting and Katzfuss, 2014]{Gneiting2014}
Gneiting, T. and Katzfuss, M. (2014).
\newblock {Probabilistic forecasting}.
\newblock {\em Annual Review of Statistics and Its Application}, 1(1):125--151.

\bibitem[Grundstr{\"o}m et~al., 2015]{grundstrom2015}
Grundstr{\"o}m, M., Hak, C., Chen, D., Hallquist, M., and Pleijel, H. (2015).
\newblock {Variation and co-variation of PM$_{10}$, particle number
  concentration, NO$_x$ and NO$_2$ in the urban air-relationships with wind
  speed, vertical temperature gradient and weather type}.
\newblock {\em Atmospheric Environment}, 120:317--327.

\bibitem[Guan et~al., 2020]{Guan2020}
Guan, Y., Johnson, M.~C., Katzfuss, M., Mannshardt, E., Messier, K.~P., Reich,
  B.~J., and Song, J.~J. (2020).
\newblock {Fine-scale spatiotemporal air pollution analysis using mobile
  monitors on Google Street View vehicles}.
\newblock {\em Journal of the American Statistical Association},
  115(531):1111--1124.

\bibitem[Guinness, 2018]{Guinness2016a}
Guinness, J. (2018).
\newblock {Permutation and grouping methods for sharpening Gaussian process
  approximations}.
\newblock {\em Technometrics}, 60(4):415--429.

\bibitem[Hastie et~al., 2017]{hastie2017extended}
Hastie, T., Tibshirani, R., and Tibshirani, R.~J. (2017).
\newblock Extended comparisons of best subset selection, forward stepwise
  selection, and the lasso.
\newblock {\em arXiv preprint arXiv:1707.08692}.

\bibitem[Heaton et~al., 2019]{Heaton2017}
Heaton, M.~J., Datta, A., Finley, A.~O., Furrer, R., Guinness, J., Guhaniyogi,
  R., Gerber, F., Gramacy, R.~B., Hammerling, D., Katzfuss, M., Lindgren, F.,
  Nychka, D.~W., Sun, F., and Zammit-Mangion, A. (2019).
\newblock {A case study competition among methods for analyzing large spatial
  data}.
\newblock {\em Journal of Agricultural, Biological, and Environmental
  Statistics}, 24(3):398--425.

\bibitem[Henderson et~al., 2007]{Henderson2007}
Henderson, S.~B., Beckerman, B., Jerrett, M., and Brauer, M. (2007).
\newblock {Application of land use regression to estimate long-term
  concentrations of traffic-related nitrogen oxides and fine particulate
  matter}.
\newblock {\em Environmental Science \& Technology}, 41(7):2422--2428.

\bibitem[Hoek et~al., 2008]{Hoek2008}
Hoek, G., Beelen, R., de~Hoogh, K., Vienneau, D., Gulliver, J., Fischer, P.,
  and Briggs, D. (2008).
\newblock {A review of land-use regression models to assess spatial variation
  of outdoor air pollution}.
\newblock {\em Atmospheric Environment}, 42(33):7561--7578.

\bibitem[Hoek et~al., 2013]{hoek2013long}
Hoek, G., Krishnan, R.~M., Beelen, R., Peters, A., Ostro, B., Brunekreef, B.,
  and Kaufman, J.~D. (2013).
\newblock Long-term air pollution exposure and cardio-respiratory mortality: a
  review.
\newblock {\em Environmental Health}, 12(1):43.

\bibitem[Holcomb et~al., 2018]{Holcomb2018}
Holcomb, D., Messier, K., Serre, M., Rowny, J., and Stewart, J. (2018).
\newblock {Geostatistical prediction of microbial water quality throughout a
  stream network using meteorology, land cover, and spatiotemporal
  autocorrelation}.
\newblock {\em Environmental Science \& Technology}, 52(14).

\bibitem[Homer et~al., 2015]{homer2015}
Homer, C., Dewitz, J., Yang, L., Jin, S., Danielson, P., Xian, G., Coulston,
  J., Herold, N., Wickham, J., and Megown, K. (2015).
\newblock Completion of the 2011 national land cover database for the
  conterminous united states--representing a decade of land cover change
  information.
\newblock {\em Photogrammetric Engineering \& Remote Sensing}, 81(5):345--354.

\bibitem[Hughes and Haran, 2013]{Hughes2013}
Hughes, J. and Haran, M. (2013).
\newblock {Dimension reduction and alleviation of confounding for spatial
  generalized linear mixed models}.
\newblock {\em Journal of the Royal Statistical Society, Series B},
  75(1):139--159.

\bibitem[Jurek and Katzfuss, 2018]{Jurek2018}
Jurek, M. and Katzfuss, M. (2018).
\newblock {Multi-resolution filters for massive spatio-temporal data}.
\newblock {\em arXiv:1810.04200}.

\bibitem[Katzfuss, 2017]{Katzfuss2015}
Katzfuss, M. (2017).
\newblock {A multi-resolution approximation for massive spatial datasets}.
\newblock {\em Journal of the American Statistical Association},
  112(517):201--214.

\bibitem[Katzfuss and Gong, 2020]{Katzfuss2017b}
Katzfuss, M. and Gong, W. (2020).
\newblock {A class of multi-resolution approximations for large spatial
  datasets}.
\newblock {\em Statistica Sinica}, 30(4):2203--2226.

\bibitem[Katzfuss and Guinness, 2019]{Katzfuss2017a}
Katzfuss, M. and Guinness, J. (2019).
\newblock {A general framework for Vecchia approximations of Gaussian
  processes}.
\newblock {\em Statistical Science}, accepted.

\bibitem[Katzfuss et~al., 2020a]{Katzfuss2018}
Katzfuss, M., Guinness, J., Gong, W., and Zilber, D. (2020a).
\newblock {Vecchia approximations of Gaussian-process predictions}.
\newblock {\em Journal of Agricultural, Biological, and Environmental
  Statistics}, 25(3):383--414.

\bibitem[Katzfuss et~al., 2020b]{GPvecchia}
Katzfuss, M., Jurek, M., Zilber, D., Gong, W., Guinness, J., Zhang, J., and
  Schaefer, F. (2020b).
\newblock {\em {GPvecchia: Fast Gaussian-process inference using Vecchia
  approximations}}.
\newblock R package version 0.1.3.

\bibitem[Kerckhoffs et~al., 2019]{kerckhoffs2019}
Kerckhoffs, J., Hoek, G., Portengen, L., Brunekreef, B., and Vermeulen, R.~C.
  (2019).
\newblock Performance of prediction algorithms for modeling outdoor air
  pollution spatial surfaces.
\newblock {\em Environmental Science \& Technology}, 53(3):1413--1421.

\bibitem[Knibbs et~al., 2014]{Knibbs2014}
Knibbs, L.~D., Hewson, M.~G., Bechle, M.~J., Marshall, J.~D., and Barnett,
  A.~G. (2014).
\newblock {A national satellite-based land-use regression model for air
  pollution exposure assessment in Australia}.
\newblock {\em Environmental Research}, 135:204--211.

\bibitem[Larkin et~al., 2017]{Larkin2017}
Larkin, A., Geddes, J.~A., Martin, R.~V., Xiao, Q., Liu, Y., Marshall, J.~D.,
  Brauer, M., and Hystad, P. (2017).
\newblock Global land use regression model for nitrogen dioxide air pollution.
\newblock {\em Environmental Science \& Technology}, 51(12):6957--6964.

\bibitem[Li et~al., 2012]{li2012sensing}
Li, J.~J., Faltings, B., Saukh, O., Hasenfratz, D., and Beutel, J. (2012).
\newblock Sensing the air we breathe—the opensense zurich dataset.
\newblock In {\em Twenty-Sixth AAAI Conference on Artificial Intelligence}.

\bibitem[Li and Sudjianto, 2005]{Li2005}
Li, R. and Sudjianto, A. (2005).
\newblock {Analysis of computer experiments using penalized likelihood in
  Gaussian kriging models}.
\newblock {\em Technometrics}, 47(2):111--120.

\bibitem[Liu et~al., 2020]{Liu2020}
Liu, H., Ong, Y.-S., Shen, X., and Cai, J. (2020).
\newblock {When Gaussian process meets big data: A review of scalable GPs}.
\newblock {\em IEEE Transactions on Neural Networks and Learning Systems},
  pages 1--19.

\bibitem[Mauzerall et~al., 2005]{mauzerall2005nox}
Mauzerall, D.~L., Sultan, B., Kim, N., and Bradford, D.~F. (2005).
\newblock Nox emissions from large point sources: variability in ozone
  production, resulting health damages and economic costs.
\newblock {\em Atmospheric Environment}, 39(16):2851--2866.

\bibitem[Meinshausen et~al., 2009]{Meinshausen2009}
Meinshausen, N., Meier, L., and B{\"{u}}hlmann, P. (2009).
\newblock {p-Values for high-dimensional regression}.
\newblock {\em Journal of the American Statistical Association},
  104(488):1671--1681.

\bibitem[Messier et~al., 2014]{Messier2014}
Messier, K., Kane, E., Bolich, R., and Serre, M. (2014).
\newblock {Nitrate variability in groundwater of North Carolina using
  monitoring and private well data models}.
\newblock {\em Environmental Science \& Technology}, 48(18).

\bibitem[Messier et~al., 2012]{Messier2012}
Messier, K.~P., Akita, Y., and Serre, M.~L. (2012).
\newblock {Integrating address geocoding, land use regression, and
  spatiotemporal geostatistical estimation for groundwater
  tetrachloroethylene.}
\newblock {\em Environmental Science {\&} Technology}, 46(5):2772--80.

\bibitem[Messier et~al., 2015]{Messier2015}
Messier, K.~P., Campbell, T., Bradley, P.~J., and Serre, M.~L. (2015).
\newblock {Estimation of groundwater Radon in North Carolina using land use
  regression and Bayesian Maximum Entropy}.
\newblock {\em Environmental Science \& Technology}, 49(16):9817--9825.

\bibitem[Miller et~al., 2020]{miller2020characterizing}
Miller, D.~J., Actkinson, B., Padilla, L., Griffin, R.~J., Moore, K., Lewis, P.
  G.~T., Gardner-Frolick, R., Craft, E., Portier, C.~J., Hamburg, S.~P., and
  Alvarez, R. (2020).
\newblock Characterizing elevated urban air pollutant spatial patterns with
  mobile monitoring in houston, texas.
\newblock {\em Environmental Science \& Technology}.

\bibitem[Mills et~al., 2015]{mills2015quantitative}
Mills, I.~C., Atkinson, R.~W., Kang, S., Walton, H., and Anderson, H. (2015).
\newblock Quantitative systematic review of the associations between short-term
  exposure to nitrogen dioxide and mortality and hospital admissions.
\newblock {\em BMJ Open}, 5(5):e006946.

\bibitem[Moore et~al., 2007]{Moore2007}
Moore, D., Jerrett, M., Mack, W., and Künzli, N. (2007).
\newblock A land use regression model for predicting ambient fine particulate
  matter across los angeles, ca.
\newblock {\em Journal of Environmental Monitoring}, 9(3):246--52.

\bibitem[{NASA/METI/AIST/Japan Spacesystems, and U.S./Japan ASTER Science
  Team}, 2019]{aster2019}
{NASA/METI/AIST/Japan Spacesystems, and U.S./Japan ASTER Science Team} (2019).
\newblock Aster global digital elevation model v003 [data set].

\bibitem[Novotny et~al., 2011]{Novotny2011}
Novotny, E.~V., Bechle, M.~J., Millet, D.~B., and Marshall, J.~D. (2011).
\newblock {National satellite-based land-use regression: NO$_2$ in the United
  States}.
\newblock {\em Environmental Science \& Technology}, 45(10):4407--4414.

\bibitem[Padr{\'o}-Mart{\'\i}nez et~al., 2012]{padro2012mobile}
Padr{\'o}-Mart{\'\i}nez, L.~T., Patton, A.~P., Trull, J.~B., Zamore, W.,
  Brugge, D., and Durant, J.~L. (2012).
\newblock Mobile monitoring of particle number concentration and other
  traffic-related air pollutants in a near-highway neighborhood over the course
  of a year.
\newblock {\em Atmospheric Environment}, 61:253--264.

\bibitem[Reyes and Serre, 2014]{Reyes2014}
Reyes, J.~M. and Serre, M.~L. (2014).
\newblock {An LUR/BME framework to estimate PM$_{2.5}$ explained by on road,
  mobile and stationary sources.}
\newblock {\em Environmental Science {\&} Technology}, 48(3):1736--44.

\bibitem[Rosenlund et~al., 2009]{rosenlund2009traffic}
Rosenlund, M., Bellander, T., Nordquist, T., and Alfredsson, L. (2009).
\newblock Traffic-generated air pollution and myocardial infarction.
\newblock {\em Epidemiology}, pages 265--271.

\bibitem[Rosenlund et~al., 2006]{rosenlund2006long}
Rosenlund, M., Berglind, N., Pershagen, G., Hallqvist, J., Jonson, T., and
  Bellander, T. (2006).
\newblock Long-term exposure to urban air pollution and myocardial infarction.
\newblock {\em Epidemiology}, pages 383--390.

\bibitem[Rosenlund et~al., 2008]{rosenlund2008traffic}
Rosenlund, M., Picciotto, S., Forastiere, F., Stafoggia, M., and Perucci, C.~A.
  (2008).
\newblock Traffic-related air pollution in relation to incidence and prognosis
  of coronary heart disease.
\newblock {\em Epidemiology}, pages 121--128.

\bibitem[Ross et~al., 2013]{Ross2013}
Ross, Z., Ito, K., Johnson, S., Yee, M., Pezeshki, G., Clougherty, J.~E.,
  Savitz, D., and Matte, T. (2013).
\newblock Spatial and temporal estimation of air pollutants in new york city:
  exposure assignment for use in a birth outcomes study.
\newblock {\em Environmental Health}, 12(1):51.

\bibitem[Sampson et~al., 2013]{Sampson2013}
Sampson, P.~D., Richards, M., Szpiro, A.~A., Bergen, S., Sheppard, L., Larson,
  T.~V., and Kaufman, J.~D. (2013).
\newblock {A regionalized national universal kriging model using partial least
  squares regression for estimating annual PM$_{2.5}$ concentrations in
  epidemiology}.
\newblock {\em Atmospheric Environment}, 75:383--392.

\bibitem[Sang et~al., 2011]{Sang2011a}
Sang, H., Jun, M., and Huang, J.~Z. (2011).
\newblock {Covariance approximation for large multivariate spatial datasets
  with an application to multiple climate model errors}.
\newblock {\em Annals of Applied Statistics}, 5(4):2519--2548.

\bibitem[Sch{\"{a}}fer et~al., 2020]{Schafer2020}
Sch{\"{a}}fer, F., Katzfuss, M., and Owhadi, H. (2020).
\newblock {Sparse Cholesky factorization by Kullback-Leibler minimization}.
\newblock {\em arXiv:2004.14455}.

\bibitem[Schindler, 1988]{schindler1988effects}
Schindler, D.~W. (1988).
\newblock Effects of acid rain on freshwater ecosystems.
\newblock {\em Science}, 239(4836):149--157.

\bibitem[Seinfeld and Pandis, 2016]{seinfeld_pandis}
Seinfeld, J.~H. and Pandis, S.~N. (2016).
\newblock {\em Atmospheric chemistry and physics: from air pollution to climate
  change}.
\newblock John Wiley \& Sons.

\bibitem[Snelson and Ghahramani, 2007]{Snelson2007}
Snelson, E. and Ghahramani, Z. (2007).
\newblock {Local and global sparse Gaussian process approximations}.
\newblock In {\em Artificial Intelligence and Statistics 11 (AISTATS)}.

\bibitem[Su et~al., 2009]{Su2009}
Su, J., Jerrett, M., and Beckerman, B. (2009).
\newblock {A distance-decay variable selection strategy for land use regression
  modeling of ambient air pollution exposures.}
\newblock {\em Science of the Total Environment}, 407(12):3890--3898.

\bibitem[Tadono et~al., 2014]{tadono2014}
Tadono, T., Ishida, H., Oda, F., Naito, S., Minakawa, K., and Iwamoto, H.
  (2014).
\newblock Precise global dem generation by alos prism.
\newblock {\em ISPRS Annals of the Photogrammetry, Remote Sensing and Spatial
  Information Sciences}, 2(4):71.

\bibitem[Tang et~al., 2019]{Tang2019}
Tang, W., Zhang, L., and Banerjee, S. (2019).
\newblock {On identifiability and consistency of the nugget in Gaussian spatial
  process models}.
\newblock {\em arXiv:1908.05726}.

\bibitem[Tibshirani, 1996]{Tibshirani1996}
Tibshirani, R. (1996).
\newblock {Regression shrinkage and selection via the lasso}.
\newblock {\em Journal of the Royal Statistical Society, Series B},
  58(1):267--288.

\bibitem[{US Environmental Protection Agency}, 1999]{epa1999nitrogen}
{US Environmental Protection Agency} (1999).
\newblock {U}.{S}. {E}{P}{A} technical bulletin: Nitrogen oxides ({N}{O}$_x$),
  why and how they are controlled.
\newblock Retrieved from U.S. Environmental Protection Agency: Available from:
  \url{https://www3.epa.gov/ttncatc1/dir1/fnoxdoc.pdf }[Last accessed on 2018
  Jul 10].

\bibitem[{US Environmental Protection Agency}, 2016]{epa2016integrated}
{US Environmental Protection Agency} (2016).
\newblock Integrated science assessment for oxides of nitrogen (final report).
\newblock Technical report, EPA/600/R-15/068. US Environmental Protection
  Agency, National Center for Environmental Assessment Research, Research
  Triangle Park, NC.

\bibitem[{US Environmental Protection Agency}, 2017]{nei2017}
{US Environmental Protection Agency} (2017).
\newblock 2017 national emissions inventory.

\bibitem[{US Environmental Protection Agency}, 2019]{epa2019data}
{US Environmental Protection Agency} (2019).
\newblock Air quality system pre-generated data files.
\newblock https://www.epa.gov/outdoor-air-quality-data/download-daily-data.

\bibitem[Vecchia, 1988]{Vecchia1988}
Vecchia, A. (1988).
\newblock {Estimation and model identification for continuous spatial
  processes}.
\newblock {\em Journal of the Royal Statistical Society, Series B},
  50(2):297--312.

\bibitem[Volk et~al., 2013]{Volk2013}
Volk, H.~E., Lurmann, F., Penfold, B., Hertz-Picciotto, I., and McConnell, R.
  (2013).
\newblock {Traffic-related air pollution, particulate matter, and autism}.
\newblock {\em JAMA Psychiatry}, 70(1):71--77.

\bibitem[Weiss et~al., 2018]{weiss2018}
Weiss, D.~J., Nelson, A., Gibson, H., Temperley, W., Peedell, S., Lieber, A.,
  Hancher, M., Poyart, E., Belchior, S., Fullman, N., et~al. (2018).
\newblock A global map of travel time to cities to assess inequalities in
  accessibility in 2015.
\newblock {\em Nature}, 553(7688):333.

\bibitem[Wu et~al., 2013]{Wu2013}
Wu, H., Wang, C., and Wu, Z. (2013).
\newblock {A new shrinkage estimator for dispersion improves differential
  expression detection in RNA-seq data}.
\newblock {\em Biostatistics}, 14(2):232--243.

\bibitem[Xie et~al., 2019]{Xie2019}
Xie, Y., Xu, L., Li, J., Deng, X., Hong, Y., Kolivras, K., and Gaines, D.~N.
  (2019).
\newblock {Spatial variable selection and an application to Virginia Lyme
  disease emergence}.
\newblock {\em Journal of the American Statistical Association}, 1459.

\bibitem[Xu et~al., 2016]{Xu2016}
Xu, X., Ha, S.~U., and Basnet, R. (2016).
\newblock A review of epidemiological research on adverse neurological effects
  of exposure to ambient air pollution.
\newblock {\em Frontiers in Public Health}, 4:157.

\bibitem[Young et~al., 2016]{Young2016}
Young, M.~T., Bechle, M.~J., Sampson, P.~D., Szpiro, A.~A., Marshall, J.~D.,
  Sheppard, L., and Kaufman, J.~D. (2016).
\newblock {Satellite-Based NO$_2$ and model validation in a national prediction
  model based on universal Kriging and land-use regression}.
\newblock {\em Environmental Science \& Technology}, 50(7):3686--3694.

\bibitem[Zhang, 2004]{Zhang2004}
Zhang, H. (2004).
\newblock {Inconsistent Estimation and Asymptotically Equal Interpolations in
  Model-Based Geostatistics}.
\newblock {\em Journal of the American Statistical Association},
  99(465):250--261.

\bibitem[Zilber and Katzfuss, 2020]{Zilber2019}
Zilber, D. and Katzfuss, M. (2020).
\newblock {Vecchia-Laplace approximations of generalized Gaussian processes for
  big non-Gaussian spatial data}.
\newblock {\em Computational Statistics {\&} Data Analysis}, accepted.

\bibitem[Zou and Hastie, 2005]{Zou2005}
Zou, H. and Hastie, T. (2005).
\newblock {Regularization and variable selection via the elastic net}.
\newblock {\em Journal of the Royal Statistical Society. Series B
  (Methodological)}, 67(2):301--320.

\end{thebibliography}


\begin{thebibliography}{}

\end{thebibliography}

\newpage

\renewcommand{\thepage}{S\arabic{page}}  
\renewcommand{\thesection}{S\arabic{section}}   
\renewcommand{\thetable}{S\arabic{table}}   
\renewcommand{\thefigure}{S\arabic{figure}}
\renewcommand{\theequation}{S\arabic{equation}}

\setcounter{page}{1}
\setcounter{section}{0}
\setcounter{table}{0}
\setcounter{figure}{0}
\setcounter{equation}{0}

\section*{\LARGE Supplementary Material}

\section{Simulation with jointly varying parameters\label{Sec:Joint-Varying}}

Supplementing the simulation study in Section \ref{sec:simulation}, we also evaluated multiple scenarios where the variables are varied jointly. Note we only vary the covariance range in space. We test a 30 scenario set that is all possible combinations of:  $\tau^2 / \sigma^{2,total}$ = $\{0.1, 0.33, 0.5, 0.67, 0.9 \}$, $\sigma^{2,total}$ = $\{1, 4 \} \times \sigma^{2,true}$, $\theta_s$ = $\{30, 300, 3000 \}$ km, and $\theta_t$ $\simeq$ 30 days.

\begin{figure}[htbp]
\centering\includegraphics[width =1\linewidth]{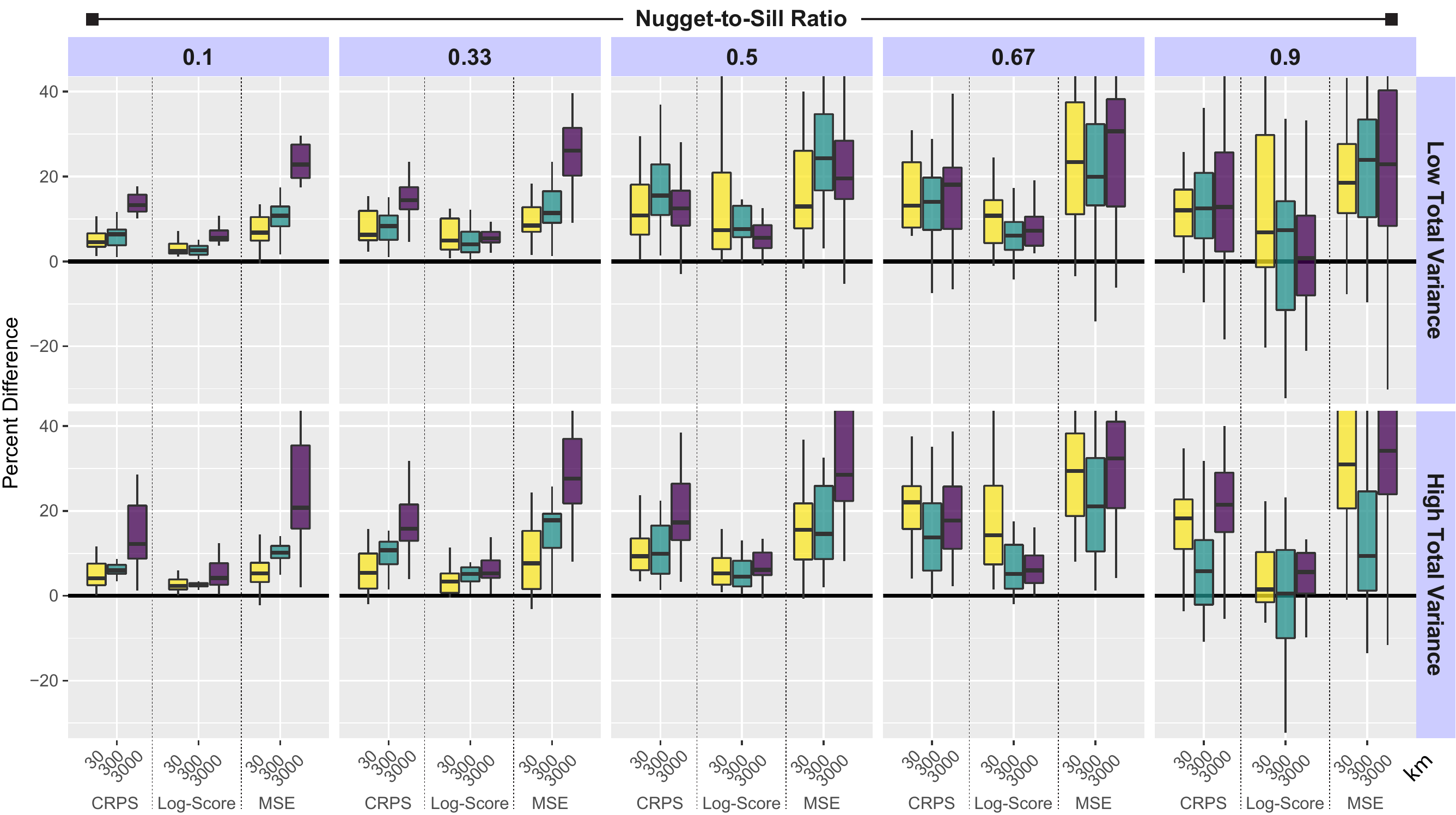}
  \caption{Percent difference between in prediction validation statistics for LURK-Local (i) and LURK-Vecchia (ii) approaches as a function of jointly varying spatiotemporal random field parameters. Percent difference is calculated as  (i-ii)/i x 100. A positive value indicates that the LURK-Vecchia performance is better. Note that some boxplots are truncated at 40 percent.}
\label{fig:Sim-ValStats-Joint}
\end{figure}

\begin{figure}[htbp]
\centering\includegraphics[width =1\linewidth]{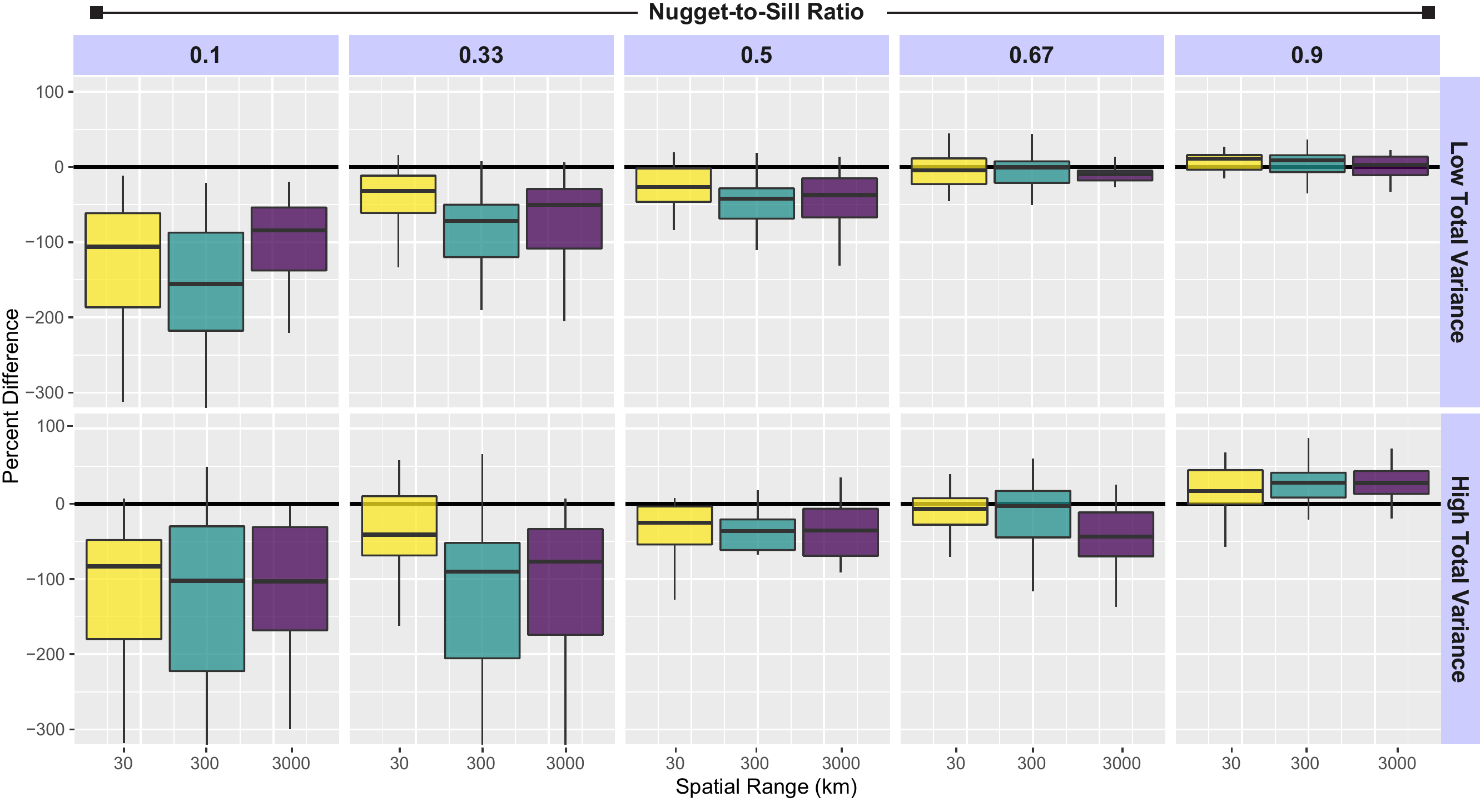}
  \caption{Percent difference between in Cohen's Kappa ($\kappa$) (overall model selection) for LURK-Local (i) and LURK-Vecchia (ii) approaches as a function of jointly varying spatiotemporal random field parameters. Percent difference is calculated as  (i-ii)/i x 100. A negative value indicates that the LURK-Vecchia performance is better.}
\label{fig:Sim-Kappa-Joint}
\end{figure}

\begin{figure}[htbp]
\centering\includegraphics[width =1\linewidth]{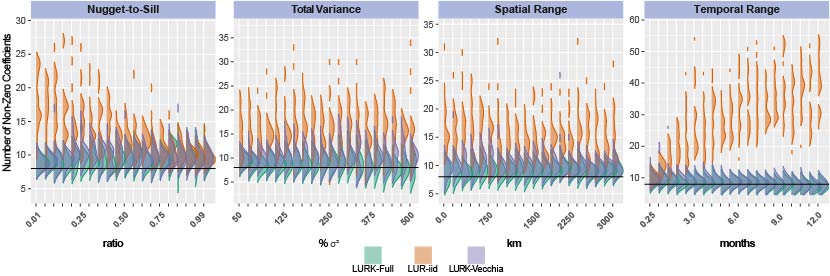}
  \caption{Ridgeline density plot comparison of the number of non-zero coefficients for simulation scenarios (\ref{sec:Approaches}) in which the spatiotemporal random field parameters are singly varied with others held constant. The horizontal black line (y=8) is the number of covariates in the true model.}
\label{fig:Sim-Betas-Nonzero-Coef}
\end{figure}

\clearpage
\section{Prediction uncertainty for NO2 application\label{sec:no2sd}}

\begin{figure}[htbp]
\centering\includegraphics[width =1\linewidth]{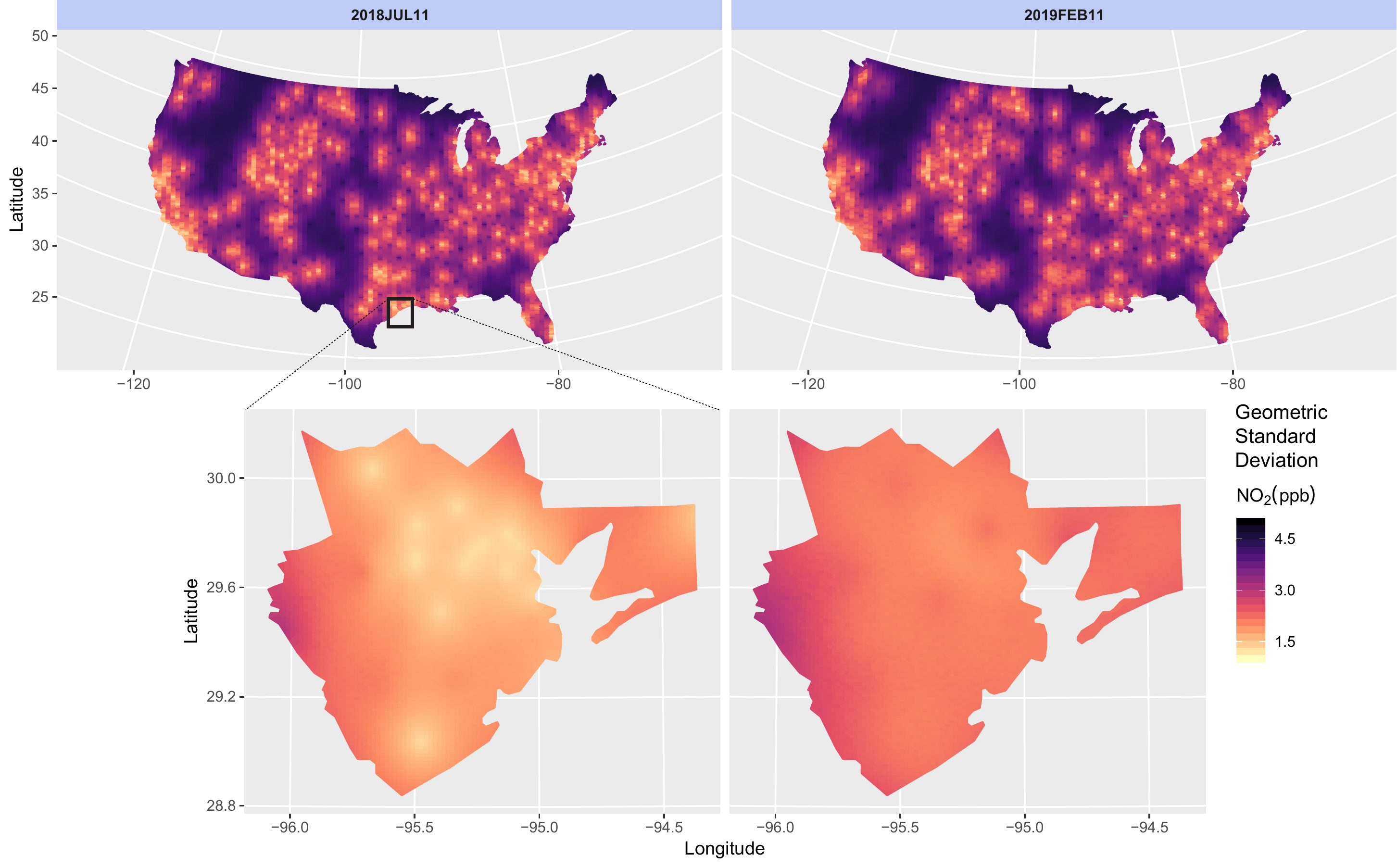}
  \caption{LURK-Vecchia prediction geometric standard deviation (exponentiated log-transform standard deviation) across the conterminous United States (top row) and the five-county subset around Houston, Texas for July 11, 2018 and February 11, 2019.}
\label{fig:VecchiaSD}
\end{figure}

Figure \ref{fig:VecchiaSD} shows prediction uncertainties for the noise-free NO$_2$ in terms of geometric standard deviations (SDs), $\exp\big(\diag(\bfSigma_P)^{1/2}\big)$. The prediction SD varies more smoothly over space than the mean in Figure \ref{fig:VecchiaPred}.

\section{Prediction and covariate comparison\label{sec:cov-comp}}

\begin{figure}
\centering\includegraphics[width =1\linewidth]{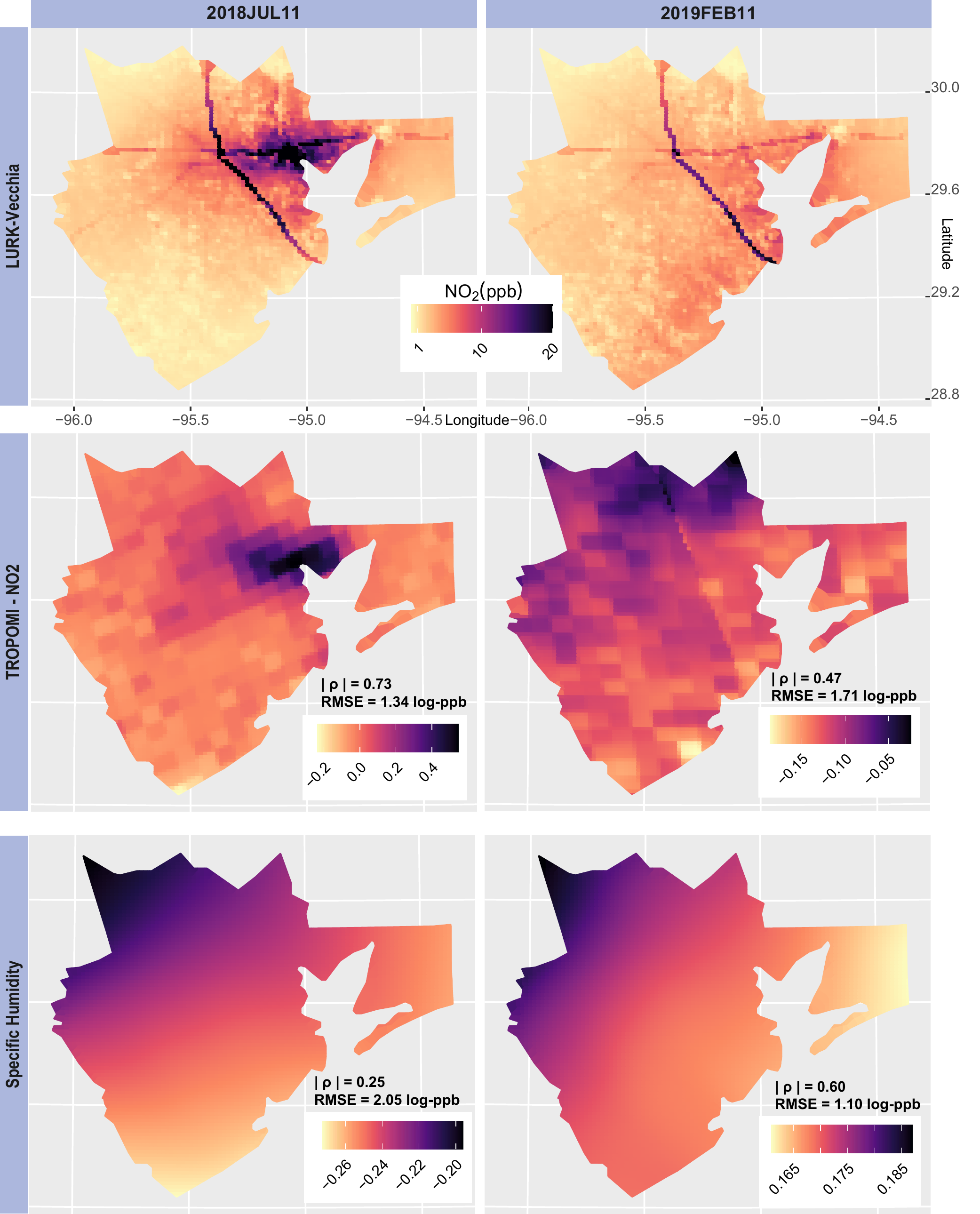}
  \caption{Comparison of the LURK-Vecchia predictions in the Houston area with predictions based only on select covariates (i.e., $\bX_{select}\hat\bfbeta_{select}$).  The rows show the (top) LURK-Vecchia final model predictions, compared to predictions using (middle) TROPOMI-NO$_2$ or (bottom) specific humidity as the only covariate(s). The left and right columns are predictions and comparisons for July 11, 2018 and February 11, 2019, respectively.}
\label{fig:Pred-Covariate-Comparison}
\end{figure}

Figure \ref{fig:Pred-Covariate-Comparison} shows point predictions (log-scale) for select covariates using only the covariate(s) and their estimated coefficients: $\bX_{select}\hat\bfbeta_{select}$. The Pearson correlation ($\rho$) and RMSE between the select covariate predictions and the respective full model prediction for that day (top row) are shown (i.e. left-middle vs left-top, left-bottom vs left-top, right-middle vs right-top, and right-bottom vs right-top). The TROPOMI-NO$_2$ covariates are clearly driving the large hot-spot of the final predictions on July 11, 2018, whereas other factors such as specific humidity have more control on February 11, 2019.

\end{document}


\maketitle


\section{Example section \label{sec:supp1}}

We can reference items such as Section \ref{sec:intro} in the main document using the xr package.

\bibliographystyle{apalike}
\bibliography{mendeley,additionalrefs}